\documentclass[lettersize,journal]{IEEEtran}
\usepackage{cite}
\usepackage{bm}
\usepackage{amsmath}
\usepackage{extarrows}
\usepackage{amssymb}
\usepackage{graphicx}
\usepackage{color}
\usepackage{enumerate}
\usepackage{bookmark} 
\graphicspath{{figure}}
\usepackage{setspace}
\usepackage{subfigure}
\usepackage{algorithm}
\usepackage{algorithmicx}
\usepackage{multirow}
\usepackage{stfloats}
\usepackage{graphbox}
\usepackage{adjustbox}
\usepackage{subcaption}
\usepackage{bbm}
\usepackage{xcolor}
\usepackage{tabularx}
\usepackage{booktabs}   
\usepackage{makecell}

\usepackage{framed}  
\usepackage{scalerel}

\newcommand{\mr}{\mathrm}

\newcommand{\BE}{\begin{equation}}
\newcommand{\EE}{\end{equation}}
\newcommand{\BS}{\begin{subequations}}
\newcommand{\ES}{\end{subequations}}
\renewcommand{\bf}{\bm}
\newcommand{\LL}[1]{\textcolor{black}{#1}}

\newcommand{\PZY}[1]{\textcolor{black}{#1}}

\newtheorem{theorem}{Theorem}

\newtheorem{assumption}{Assumption}
\newtheorem{conjecture}{Conjecture}

\newtheorem{lemma}{Lemma}
\newtheorem{corollary}{Corollary}
\newcommand{\tabincell}[2]{\begin{tabular}{@{}#1@{}}#2\end{tabular}}

\allowdisplaybreaks \allowdisplaybreaks[2]

\ifCLASSINFOpdf
\graphicspath{{figure/}}
\else

\fi
\begin{document}



\title{Achievable Rate and  Coding Principle for MIMO Multicarrier Systems With Cross-Domain MAMP Receiver Over Doubly Selective Channels}

\author{\IEEEauthorblockN{Yuhao Chi, \emph{Senior Member, IEEE}, Zhiyuan Peng, Lei Liu, \emph{Senior Member, IEEE}, \\ Ying Li, \emph{Member, IEEE},  Yao Ge, \emph{Member, IEEE},  and Chau Yuen, \emph{Fellow, IEEE}}

\thanks{Yuhao Chi, Zhiyuan Peng, and Ying Li are with the State Key Laboratory of Integrated Services Networks, School of Telecommunications Engineering, Xidian University, Xi'an, Shaanxi, 710071, China (e-mail: yhchi@xidian.edu.cn, zypeng\_1@stu.xidian.edu.cn, yli@mail.xidian.edu.cn).}

\thanks{Lei Liu is with the Zhejiang Provincial Key Laboratory of Information Processing, Communication and Networking, College of Information Science and Electronic Engineering, Zhejiang University, Hangzhou, Zhejiang, 310007, China, and also with the State Key Laboratory of Integrated Services Networks, Xidian University, Xi’an, Shaanxi, 710071, China (e-mail: lei\_liu@zju.edu.cn).}

\thanks{Yao Ge is with the AUMOVIO-NTU Corporate Lab, Nanyang Technological University, Singapore 639798 (e-mail:yao.ge@ntu.edu.sg).}

\thanks{Chau Yuen is with the School of Electrical and Electronics Engineering, Nanyang Technological University, Singapore 639798 (e-mail: chau.yuen@ntu.edu.sg).}
}

\maketitle
\begin{abstract}
The integration of multicarrier modulation and multiple-input-multiple-output (MIMO) is  critical for reliable transmission of wireless signals in complex environments, which significantly improve spectrum efficiency. Existing studies have shown that popular orthogonal time frequency space (OTFS) and affine frequency division multiplexing (AFDM) offer significant advantages over orthogonal frequency division multiplexing (OFDM) in uncoded doubly selective channels. However, it remains uncertain whether these benefits extend to coded systems.
Meanwhile, the information-theoretic limit analysis of coded MIMO multicarrier systems and the corresponding low-complexity receiver design remain unclear.
To overcome these challenges, this paper proposes a multi-slot cross-domain memory approximate message passing (MS-CD-MAMP) receiver as well as  develops its information-theoretic (i.e., achievable rate) limit 
and optimal coding principle for MIMO-multicarrier modulation (e.g., OFDM, OTFS, and AFDM) systems.  The proposed MS-CD-MAMP receiver can exploit not only the time domain channel sparsity for low complexity but also the corresponding symbol domain constellation constraints for performance enhancement. \LL{Meanwhile, limited by the high-dimensional complex state evolution (SE), a simplified single-input single-output variational SE is proposed to derive the achievable rate of MS-CD-MAMP and the optimal coding principle with the goal of maximizing the achievable rate.} Numerical results show that coded MIMO-OFDM/OTFS/AFDM with MS-CD-MAMP achieve the same maximum achievable rate in doubly selective channels, whose finite-length performance with practical optimized low-density parity-check (LDPC) codes is only $0.5\sim1.8$~dB away from the associated theoretical limit, and has $0.8\sim 4.4$~dB gain over the well-designed point-to-point LDPC codes.
\end{abstract}


\section{Introduction}
\subsection{Overview of Multicarrier Modulations}
{With the rapid expansion of wireless networks and IoT, B5G and 6G communications are expected to support reliable connectivity in high-mobility scenarios, such as high-speed railways\cite{OTFS-TSMA}, unmanned aerial vehicles (UAVs) \cite{CJE_UAV}, and low-earth orbit (LEO) satellites \cite{CJE_LEO}. However, in current 5G systems, orthogonal frequency-division multiplexing (OFDM) focuses on overcoming inter-symbol interference in static multipath channels\cite{tse2005fundamentals}. It struggles to handle the inter-carrier interference caused by Doppler spread in high-mobility channels. Therefore, designing multicarrier modulation schemes that can effectively support both static and high-mobility scenarios is a key enabler for future B5G and 6G systems.}




In recent years, orthogonal time frequency space (OTFS) is proposed as an emerging modulation technique to ensure reliable high-mobility wireless communications \cite{OTFS,OTFS_Survey}. Unlike OFDM, digitally modulated signals in OTFS are placed in the two-dimensional (2D) delay-Doppler (DD) domain and then mapped to the time-frequency domain, thereby exploiting both delay and Doppler diversity for significant performance gains. {To reduce implementation complexity, a kind of modulation based on the sparse Walsh-Hadamard transform is proposed in~\cite{WHM} to achieve lower complexity by replacing the symplectic finite Fourier transform (SFFT) in OTFS and preserving the similar error performance.} Independently, affine frequency division multiplexing (AFDM) \cite{AFDM_twc} achieves notable diversity gains using one-dimensional (1D) orthogonal chirp sequences, which distinguish multipath signals through adjustable chirp parameters. Fundamentally, both OTFS and AFDM aim to maximize diversity by leveraging modulation matrices to separate signals in delay and Doppler dimensions. Their effective channel matrices are sparse, enabling the design of low-complexity detection algorithms. More recently, interleave frequency division multiplexing (IFDM) is proposed in \cite{IFDM}, differing from OTFS and AFDM in that its goal is to achieve the replica channel capacity and replica \emph{maximum a posteriori} (MAP) optimality. In IFDM, the effective channel matrix is randomly dense with statistically smooth properties, thanks to randomly interleaved inverse Fourier transform. Consequently, all signals are randomly and independently assigned to the whole subcarriers, ensuring that all signals undergo statistical channel fading sufficiently. {Furthermore, multiple-input multiple-output (MIMO) can be combined with OTFS, AFDM, and IFDM to further improve spectral efficiency and transmission reliability for high mobility scenarios~\cite{MIMO-OTFS-LMMSE,MIMO-OTFS-Dect,MIMO-OTFS-WCL,MIMO-AFDM-CE,IFDM}}. However, in MIMO-multicarrier scenarios, the system dimensions increase and the interference becomes more severe, as additional inter-antenna interference is introduced, posing new challenges to transceiver design. Meanwhile, most of the work on MIMO-multicarrier systems focuses solely on uncoded systems, often neglecting the advantages of channel coding.

\subsection{Performance Analysis for Coded Multicarrier Systems}
To date, a few works have been conducted on the performance analysis of coded multicarrier systems. In \cite{LDPCOFDM}, irregular low-density parity-check (LDPC) codes are optimized using a turbo iterative receiver for MIMO-OFDM over frequency-selective channels. The pairwise-error probability (PEP) of coded MIMO-OFDM is studied for different signal constellations in the frequency-selective channels\cite{PEPCodedOFDM}. However, the turbo iterative receiver is strictly suboptimal for discrete input signaling\cite{LeiTIT2021}, and the study of coded MIMO-OFDM systems in \cite{LDPCOFDM,PEPCodedOFDM} is limited to frequency-selective fading, without taking double-selective channels into account. 

Recently, a conditional PEP is derived as the performance upper bound of coded OTFS systems \cite{codedOTFS}, revealing a trade-off between coding gain and diversity gain. Similarly, the conditional PEP of coded AFDM systems is presented in \cite{yin2023error}, showing the similar trade-off. Numerical results in \cite{codedOTFS,yin2023error} reveal that the coded OTFS/AFDM significantly outperforms the coded OFDM. However, these analyses rely on a strong assumption: the encoded codeword length equals the modulation dimension, i.e., a single OTFS or AFDM symbol transmission slot. This assumption is impractical for practical systems with large data payloads, where long codewords must be split into multiple blocks spanning several transmission slots. Additionally, \cite{codedOTFS,yin2023error} only analyze the impact of diversity gain and do not clarify the information-theoretic limits of the system.  {Furthermore, in \cite{codedOTFS_OFDM}, performance comparisons between coded OTFS and OFDM using iterative minimum mean-square error 
(MMSE) and message passing (MP) receivers under discrete constellations (e.g., quadrature phase-shift keying (QPSK), 16-quadrature amplitude modulation (16QAM)), which shows OTFS outperforms OFDM with QPSK signaling, whereas OFDM achieves better performance with 16QAM signaling. However, iterative MMSE and MP receivers are suboptimal for discrete input signaling, as discussed in the following subsection. Moreover, short-length LDPC codes are used, which may lead to significant performance degradation—especially under high-order constellations.}  Therefore, the information-theoretic limits and optimal coding design of coded MIMO multicarrier systems remain open issues. Equally important is the development of practical receivers capable of approaching or achieving these limits.


\subsection{Advanced Detection Technology for Multicarrier Systems}
Detection techniques in multicarrier modulation are crucial, as they directly determines the communication performance of practical systems. Since the mathematical model of multicarrier systems can be formulated as a standard linear model, most widely used detection algorithms are applicable across different schemes. Linear detectors, such as linear minimum mean-square error (LMMSE) and zero forcing (ZF), can be employed in OFDM, OTFS, and AFDM systems \cite{LMMSEOFDM,LMMSE&ZF,AfdmLow}. Detection complexity can be further reduced by exploiting the structure of the channel matrices. For instance, in OTFS systems with finite frame duration and fractional Doppler, direct application of LMMSE or ZF requires matrix inversion, incurring high complexity. To mitigate this, a low-complexity iterative successive interference cancellation-based MMSE (SIC-MMSE) detector was proposed in \cite{SIC-MMSE}, which sequentially performs MMSE estimation on sparse time-domain channel matrix columns while reusing filter weights, thereby greatly reducing inversion overhead. Nevertheless, linear detectors cannot exploit the a priori distribution of transmitted signals, limiting their ability to achieve reliable signal recovery.
To address this issue, low-complexity Gaussian message-passing (GMP) and expectation propagation (EP) algorithms are developed as iterative detectors for OTFS \cite{GMP2018}\cite{GeYao2021}. However, due to numerous short loops in the equivalent factor graph of OTFS, GMP and EP detectors may diverge, requiring careful damping parameter tuning. {An iterative LMMSE parallel successive interference cancellation (PIC) equalizer is developed based on classical Turbo principle\cite{LMMSE-PIC}, which is strictly suboptimal for discrete input signaling \cite{MaTWC2019,LeiOptOAMP,YuhaoTcom2022,CodeMAMP2023}.} Therefore, cross-domain detectors have been proposed that employ linear detection in the time domain and MMSE demodulation in the symbol domain\footnote{\LL{The ``symbol domain" denotes the constellation symbol mapping under a given modulation scheme, such as the frequency domain in OFDM, the delay-Doppler domain in OTFS, and the affine frequency domain in AFDM.}}, effectively reducing the complexity of signal detection by leveraging the sparsity of the time-domain channel\footnote{\LL{The ``time-domain channel" refers to the ``time-delay domain channel". Following the terminology in the existing book \cite{das2022orthogonal}, particularly for cross-domain receiver works \cite{OTFS-OAMP,IFDM}, we use ``time-domain channel" as a shorthand for ``time-delay domain channel" in this paper.}}. For instance, in \cite{OTFS-OAMP}, a robust cross-domain orthogonal approximate message passing (CD-OAMP) detector is developed, which iteratively operates between the linear detection (LD) in time domain and nonlinear detection (NLD) in the DD domain. Similarly, a DD-OAMP is developed for OTFS in the DD domain \cite{DD-OAMP,OAMP/VAMP}. Nevertheless, since the LD in CD/DD-OAMP is based on LMMSE and limited by matrix inversion, it fails to effectively leverage channel sparsity, resulting in high complexity for large-scale systems. To tackle this challenge, a DD-domain memory approximate message passing (DD-MAMP) detector is proposed in \cite{MAMPOTFSconf}, utilizing a memory matched filter (MF) to exploit the sparsity of DD channels. However, DD-MAMP overlooks the even sparser time-domain channels. Recently, a cross-domain MAMP (CD-MAMP) detector for IFDM is proposed in \cite{IFDM}, which achieves extremely low complexity by using a memory MF in more sparse time-domain channel matrices. Although signal detection algorithms can be extended directly to MIMO, they are currently focused on uncoded systems, ignoring the importance of channel coding and decoding in practical systems. Consequently, the joint design of receivers for coded systems remains an open problem.

\subsection{Contributions}
In this paper, we present a low-complexity and high-reliability multi-slot cross-domain MAMP (MS-CD-MAMP) receiver, as well as its information-theoretic limit (i.e., maximum achievable rate) analysis and optimal coding principle for MIMO-multicarrier modulations (i.e., OFDM, OTFS, and AFDM). Specifically, 
the MS-CD-MAMP receiver fully exploits the sparsity of time-domain channel matrices by employing multiple memory-based MFs across multiple time slots for signal estimation, while jointly performing signal demodulation and channel decoding in the corresponding symbol domain. Each memory MF relies on all prior iteration estimates, and the eigenvalue distributions of channel matrices vary across time slots, leading to a complicated multidimensional state evolution (SE) that is difficult to analyze. To address this difficulty, a simplified single-input-single-output (SISO) variational state evolution (VSE) is derived by leveraging the SE fixed-point consistency of MS-CD-MAMP and MS-CD-OAMP, along with the multi-slot VSE averaging strategy. Based on the SISO VSE, the achievable rates of different modulations with MS-CD-MAMP are derived using the I-MMSE lemma\cite{GuoTIT2005}, while the optimal coding principles are derived with the aim of maximizing the achievable rates. Furthermore, a kind of practical LDPC code is developed for different modulations with correlated MIMO channels and different system parameters. The main contributions of this paper are summarized as follows:
\begin{itemize}
\item {A low-complexity high-reliability MS-CD-MAMP receiver is proposed for coded MIMO-multicarrier modulation systems, with performance comparable to the practical state-of-the-art OAMP receiver.}
\item {The multi-slot SE is proposed to demonstrate the mean-square error equivalence of cross-domain MAMP/OAMP and DD-domain MAMP/OAMP.}
\item {\LL{A simplified SISO VSE is proposed for MS-CD-MAMP to analyze the achievable rate and establish the optimal coding principle.} On this basis, coded MIMO-OFDM, MIMO-OTFS, and MIMO-AFDM with MS-CD-MAMP and optimal coding are shown to achieve the same maximum achievable rates. Meanwhile, the achievable rates analysis are compared under various parameter configurations and arbitrary input distributions.}
\item {A kind of practical LDPC code is optimized for MIMO-OFDM/OTFS/AFDM, whose theoretical decoding thresholds are about $0.2$~dB away from the associated limits. Numerical results show that the MIMO-OFDM/OTFS/AFDM with MS-CD-MAMP and optimized LDPC codes can achieve almost the same bit error rate (BER) performances and $0.8\sim4.4$~dB gains over those with MS-CD-MAMP and well-designed point-to-point (P2P) LDPC codes.}
\end{itemize}

Part of the results in this paper has been published in \cite{chiICC2025}. In this paper, we additionally provide the derivation of the achievable rates, detailed proofs, and more numerical results.

\subsection{Notations}
Matrix symbols (column vectors) are bold uppercase (lowercase) letters.
The transpose, conjugate transpose, and inverse operations are shown by the notations $[\cdot]^{\rm{T}}$, $[\cdot]^{\rm{H}}$, and $[\cdot]^{-1}$, respectively. 
$\bf{I}$ and $\bf{0}$ are identity matrix and zero matrix or vector. 
The minimal value, maximum value, and cardinality of the set $\mathcal{S}$ are indicated, respectively, by the variables $\rm min(\mathcal{S})$, $\rm max(\mathcal{S})$, and $|\mathcal{S}|$. 
Denote $\|\bf{a}\|$ for vector's $\ell_2$-norm $\bf{a}$, $\mr{tr}(\bf{A})$ for the trace of matrix $\bf{A}$, $\mathcal{CN}(\bf{\mu},\bf{\Sigma})$ for the circularly-symmetric Gaussian distributions with mean $\bf{\mu}$ and covariance $\bf{\Sigma}$, and $\left \langle \bf{A}_{M \times N} \mid \bf{B}_{M \times N} \right \rangle \equiv \tfrac{1}{N}\bf{A}^{\rm{H}}_{M \times N}\bf{B}_{M \times N}$, $\mr{E}\{a|b\}$ for the expectation of $a$ conditional on $b$, $\rm{mmse}\{a|b\}$ for $\mr{E}\{(a-E\{a|b\})^2|b\}$. 
$X \sim Y$ represents that $X$ follows the distribution $Y$.

\subsection{Paper Outline}
This paper is organized as follows. Section II presents the system model and key challenges of coded MIMO-multicarrier systems. The MS-CD-MAMP receiver and state evolution are provided in Section III. The corresponding achievable rates analysis and coding principle are presented for MIMO-OFDM/OTFS/AFDM in Section IV. Numerical simulations and the conclusion are presented in Sections V and VI.


\section{System Model and Key Challenges}
In this section, we present a coded MIMO-multicarrier system model and review OFDM, OTFS, and AFDM modulations. Then, the existing key challenges are discussed. 

\begin{figure*}[t!]\vspace{-0.4cm}
	\centering 
	\includegraphics[width=1\textwidth]{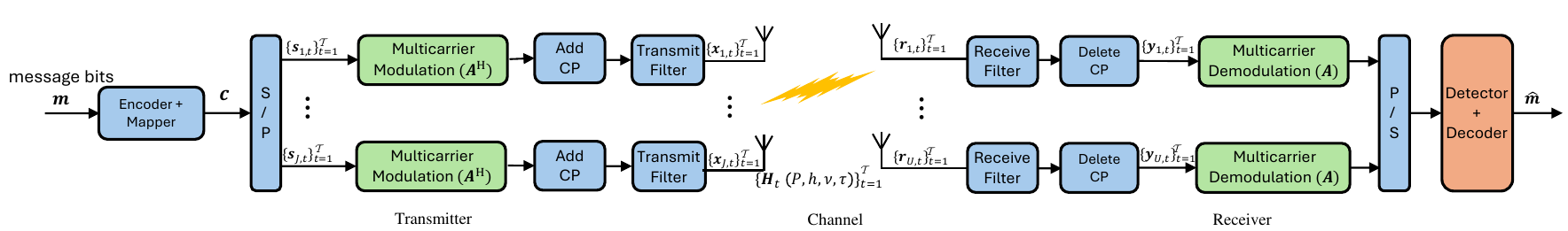}
	\caption{{Block diagram of a coded MIMO-multicarrier system}.}\label{Fig:systemModel}  
\end{figure*}
\subsection{MIMO-Multicarrier System Model}\label{sec:MIMOsys}
Fig.~\ref{Fig:systemModel} shows a coded MIMO-multicarrier modulation system with a $J$-antenna transmitter and an $U$-antenna receiver. At the transmitter, a message bit vector $\bf{m}$ is encoded and mapped to $\bf{c}\in \mathcal{S}^{M \times 1}$, 
which is converted to $\{\{\bf{s}_{1,t}\}_{t=1}^{\mathcal{T}}, ..., \{\bf{s}_{J,t}\}_{t=1}^{\mathcal{T}}\}$, where 
$\bf{s}_{j, t} \in \mathcal{S}^{N \times 1}$ denotes the transmitted signal of the $j$-th antenna at the $t$-th time slot, $\mathcal{S}$ denotes the constellation set,  $j=1,..., J$, and $t=1, ..., \mathcal{T}$.
For simplicity, we assume $M=NJ\mathcal{T}$, where $\mathcal{T}$ is the total number of transmission time slots. Followed by a specific multicarrier modulation, the time-domain signal is generated as $\bf{x}_{j,t}$, i.e., 
\BE\label{Eqn:mcsignal}
\bf{x}_{j,t}=\bf{A}^{\rm{H}}\bf{s}_{j,t},
\EE
where $\bf{A}$ denotes the $N$-point multicarrier transform matrix.
{After adding the CP and passing through the transmit filter, $\bf{x}_{j,t}$ is transmitted over the $j$-th antenna at the $t$-th slot.}


{
The received time domain signal first enters a received filter. After discarding the CP, the received signal ${y}_{u,t}[n]$ at the $u$-th antenna in the $t$-th time slot is given by
\BE\label{Eqn:rev_u}
y_{u,t}[n]=\sum\limits_{j=1}^{J} \sum\limits_{\iota=0}^{\mathcal{I}_{uj}-1} h_{uj,t}[n,\iota]x_{j,t}[{[n-\iota]}_{N}]+w_{u,t}[n],
\EE
where $[a]_N$ denotes $a$ mod $N$, $u=1, ..., U$, $n=0, ..., N-1$, $t=1, ..., \mathcal{T}$,
$\mathcal{I}_{uj}$ denotes the maximal number of channel taps between the $j$-th transmit antenna and the $u$-th receive antenna, and ${w}_{u,t}[n] \sim \mathcal{CN}({0},\sigma^2)$ is an additive white Gaussian noise (AWGN) variable. The channel impulse response $h_{uj,t}[n, \iota]$ between the $j$-th transmit antenna and the $u$-th receive antenna is given by
\BE\label{Eqn:chl}
\!\!h_{uj,t}[n, \iota] \!= \!\sum\limits_{i=1}^{P_{uj}} h_{uj,t}^i e^{j2\pi \nu_{uj,t}^i(nT_s -\iota T_s)}\mr{P_{rc}}(\iota T_s-\tau_{uj,t}^i),
\EE
where {$P_{uj}$} is the maximal number of multipaths between the $j$-th transmit antenna and the $u$-th receive antenna, $T_s$ represents the system sampling interval. The notations $h_{uj,t}^i$, $\tau_{uj,t}^i$, and $\nu_{uj,t}^i$ denote the channel gain, delay,  and Doppler shift associated with $i$-th path at the $t$-th slot, respectively. 
$\mr{P_{rc}}(\cdot)$ is the overall raised-cosine rolloff filter when the practical root raised-cosine (RRC) pulse shaping filters are employed at the transceiver to control signal bandwidth and reject out-of-band emissions.}

Meanwhile, in practical MIMO channels, $h_{uj,t}^i$ is also related to antenna correlations, which depend on the propagation environment, antenna pattern, and relative positions of the transmit and receive antennas. {Assuming that the transmitter and receiver are sufficiently separated, the correlation matrices $\bf{R}_{\rm tx}\in\mathbb{C}^{J \times J}$ and $\bf{R}_{\rm rx}\in \mathbb{C}^{U \times U}$ characterize the correlations among the sub-channels at the transmitter and the receiver, respectively. which are widely adopted in \cite{MIMO-OTFS-WCL,Channel2} and their references therein. This model enables the investigation of the impact of inter-antenna
correlation on the information-theoretic performance limits of the MIMO multicarrier systems.} The elements of $\bf{R}_{\rm tx}$ and $\bf{R}_{\rm rx}$ are 
\BE\label{Eqn:TRx}
\begin{aligned}
&\bf{R}_{\rm{tx}}[l, k]=\left\{\begin{array}{ll}
\rho_{\rm{tx}}^{l-k}, & k \leq l \\
\left(\rho_{\rm{tx}}^{k-l}\right)^*, & k>l
\end{array}, k, l \in\left\{1, \ldots, J\right\}, \right.  \\
&\bf{R}_{\rm{rx}}[l, k]=\left\{\begin{array}{ll}
\rho_{\rm{rx}}^{l-k}, & k \leq l \\
\left(\rho_{\rm{rx}}^{k-l}\right)^*, & k>l
\end{array}, k, l \in\left\{1, \ldots, U\right\}, \right.
\end{aligned}
\EE
where $\rho_{\rm{tx}}$, $\rho_{\rm{rx}} \in[0,1)$ denote the correlation level at $\bf{R}_{\rm tx}$ and $\bf{R}_{\rm rx}$, respectively. {As stated in \cite{Channel2}, although this correlation MIMO model in \eqref{Eqn:TRx} may  not be an accurate model for some real-world scenarios, it is a simple single-parameter model that allows one to study the effect of correlation on the MIMO capacity in an explicit way and to get some insight. Meanwhile, this correlation model in \eqref{Eqn:TRx} is physically reasonable in the sense that the correlation decreases with increasing distance between receive antennas and it also corresponds to some realistic physical  configurations (e.g., \cite[p.26]{turin1962optimal}). Comparison with the measurement results \cite{MIMO_channel} also shows that it provides reasonable conclusions when applied to MIMO systems. It should be noted that the receiver design and analysis in this paper do not directly rely on the correlation model in \eqref{Eqn:TRx}, but rather utilize the time-domain channel matrix 
$\bf{H}_t$. As a result, our study is not limited to this specific correlation model, and channel modeling itself is not the main focus of this work. If other correlation channel models are considered, the proposed scheme can be directly extended accordingly.}

\LL{Similar to MIMO-OTFS channel model in \cite{MIMO-OTFS-chanl}, let $\bf{\mathcal{H}}_{i,t}$ be an $U \times J$ channel gain matrix associated with the $i$-th path in the $t$-th slot, whose entries obey IID complex Gaussian distribution, i.e., $\bf{\mathcal{H}}_{i,t}[u, j] \sim \mathcal{C} \mathcal{N}\left(0, \sigma_{i,t}^2(u, j)\right)$ with the average power $\sigma_{i,t}^2(u, j)$. Then, the spatially correlated channel gain matrix is obtained by
\BE\label{Eqn:correlatedChannelMatrix}
\bar{\bf{\mathcal{H}}}_{i,t}=\bf{C}_{\rm{rx}} \cdot \bf{\mathcal{H}}_{i,t} \cdot \bf{C}_{\rm{tx}}^{\rm H},
\EE
where $\bf{C}_{\rm{tx}}\in\mathbb{C}^{J \times J}$ and $\bf{C}_{\rm{rx}}\in\mathbb{C}^{U \times U}$ are correlation-shaping matrices obtained by the Cholesky decomposition of $\bf{R}_{\rm{tx}}$ and $\bf{R}_{\rm{rx}}$, that is, $\bf{R}_{\rm{tx}}=\bf{C}_{\rm{tx}} \cdot \bf{C}_{\rm{tx}}^{\rm{H}}$ and $\bf{R}_{\rm{rx}}=\bf{C}_{\rm{rx}} \cdot \bf{C}_{\rm{rx}}^{\rm{H}}$. As a result, let $h_{uj,t}^i=\bar{\bf{\mathcal{H}}}_{i,t}[u, j]$ in \eqref{Eqn:chl}. 
}

Based on \eqref{Eqn:chl}, the received signal in \eqref{Eqn:rev_u} can be represented in matrix form as 
\BE\label{Eqn:Rev_y}
\bf{y}_{u,t} =\textstyle\sum\nolimits_{j=1}^{J}\bf{H}_{uj,t}\bf{x}_{j,t} + \bf{w}_{u,t} = \bf{H}_{u,t} \bf{x}_t +\bf{w}_{u,t},
\EE
where $\bf{H}_{uj,t}$ is given in \eqref{eqn_dbl_x}, $\bf{H}_{u,t} = [\bf{H}_{u1,t}, ..., \bf{H}_{uJ,t}] \in \mathbb{C}^{N\times NJ}$, $\bf{x}_t=[\bf{x}_{1,t}^{\rm{T}}$, $...,\bf{x}_{J,t }^{\rm{T}}]^{\rm{T}} \in \mathbb{C}^{NJ \times 1}$, and $\bf{w}_{u,t} \in \mathbb{C}^{N\times 1}\sim \mathcal{CN}(\bf{0}, \sigma^2\bf{I})$.


\begin{figure*}[!b]
\tiny
\hrulefill
\begin{equation}
\label{eqn_dbl_x}
\bf{H}_{uj,t}=\\
 \left[\begin{array}{cccccccc}h_{uj,t}[0,0] & 0 & \cdots & 0 & h_{uj,t}[0, \mathcal{I}_{uj}-1] & h_{uj,t}[0, \mathcal{I}_{uj}-2] & \cdots & h_{uj,t}[0,1] \\ h_{uj,t}[1,1] & h_{uj,t}[1,0] & 0 & \cdots & 0 & h_{uj,t}[1, \mathcal{I}_{uj}-1] & \cdots & h_{uj,t}[1,2] \\ \vdots & \ddots  & \ddots & \ddots & \ddots & \ddots & \ddots & \vdots \\ h_{uj,t}[\mathcal{I}_{uj}-1, \mathcal{I}_{uj}-1] & \cdots & \cdots & h_{uj,t}[\mathcal{I}_{uj}-1, 1] & h_{uj,t}[\mathcal{I}_{uj}-1, 0] & 0 & \cdots & 0 \\ \vdots & \ddots & \ddots & \ddots & \ddots & \ddots & \ddots & \vdots \\ 0 & \cdots & 0 & h_{uj,t}[N-1, \mathcal{I}_{uj}-1] &\cdots & \cdots & h_{uj,t}[N-1,1] & h_{uj,t}[N-1,0]\end{array}\right]
\end{equation}
\end{figure*}



 \begin{figure*}[t]
\centering
\subfigure[Time-domain $\bf{H}_t$]{
\begin{minipage}[t]{0.24\textwidth}
\centering
\includegraphics[width=0.9\textwidth]{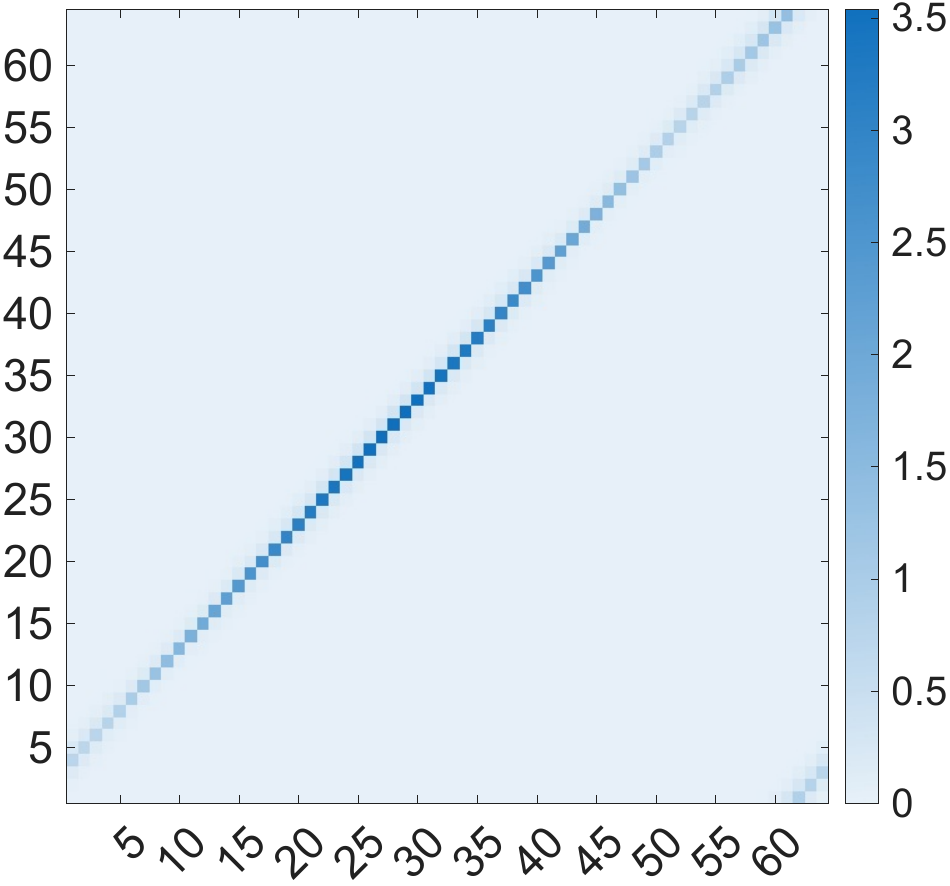}
\end{minipage}%
}%
\subfigure[$\bf{H}_{{\rm{eff}},t}^{\text{OFDM}}$]{
\begin{minipage}[t]{0.24\textwidth}
\centering
\includegraphics[width=0.9\textwidth]{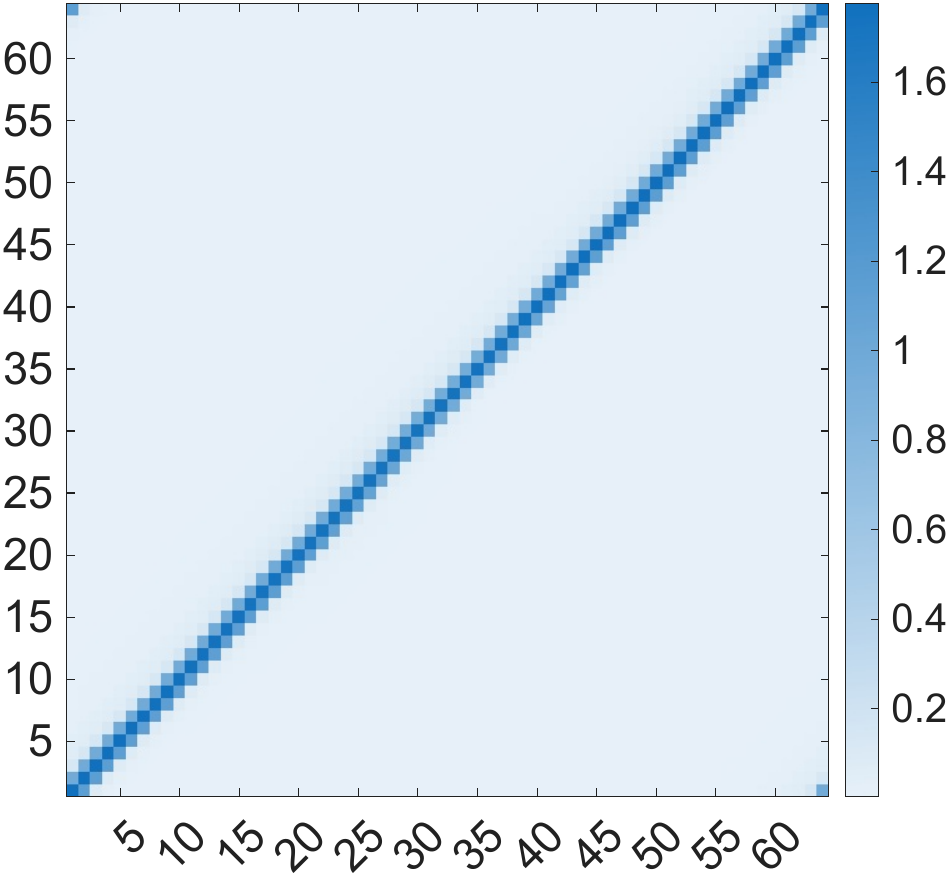}
\end{minipage}%
}%
\subfigure[$\bf{H}_{{\rm{eff}},t}^{\text{OTFS}}$]{
\begin{minipage}[t]{0.24\textwidth}
\centering
\includegraphics[width=0.9\textwidth]{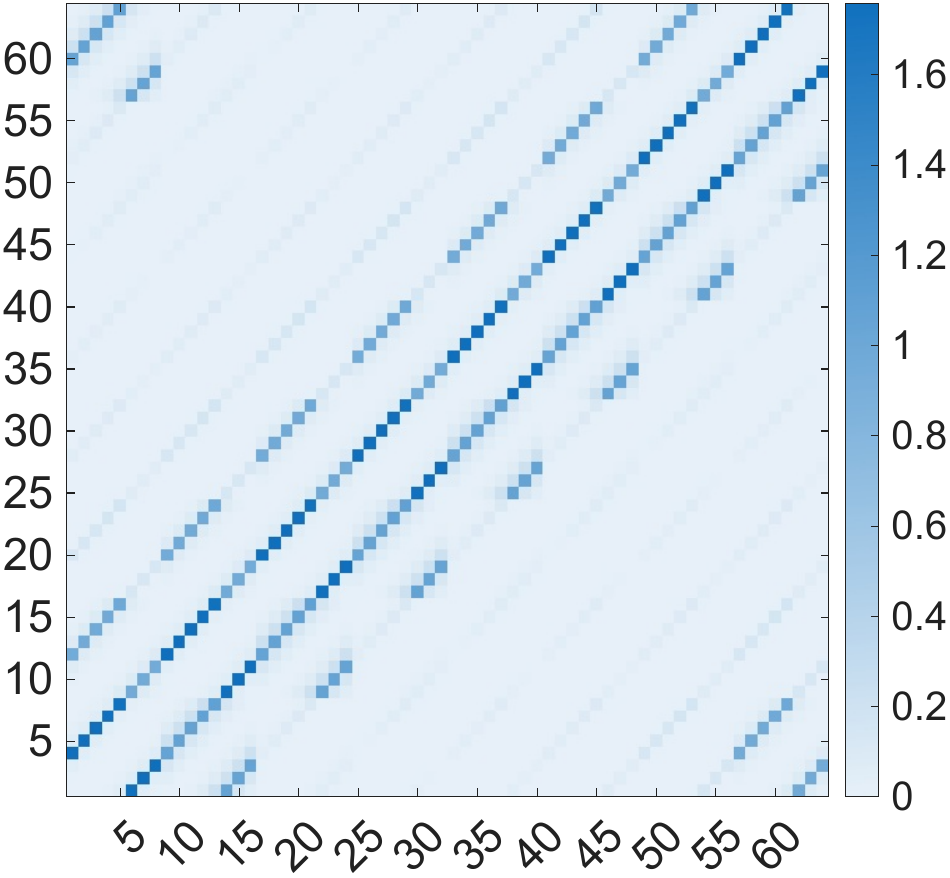}
\end{minipage}
}%
\subfigure[$\bf{H}_{{\rm{eff}},t}^{\text{AFDM}}$]{
\begin{minipage}[t]{0.24\textwidth}
\centering
\includegraphics[width=0.9\textwidth]{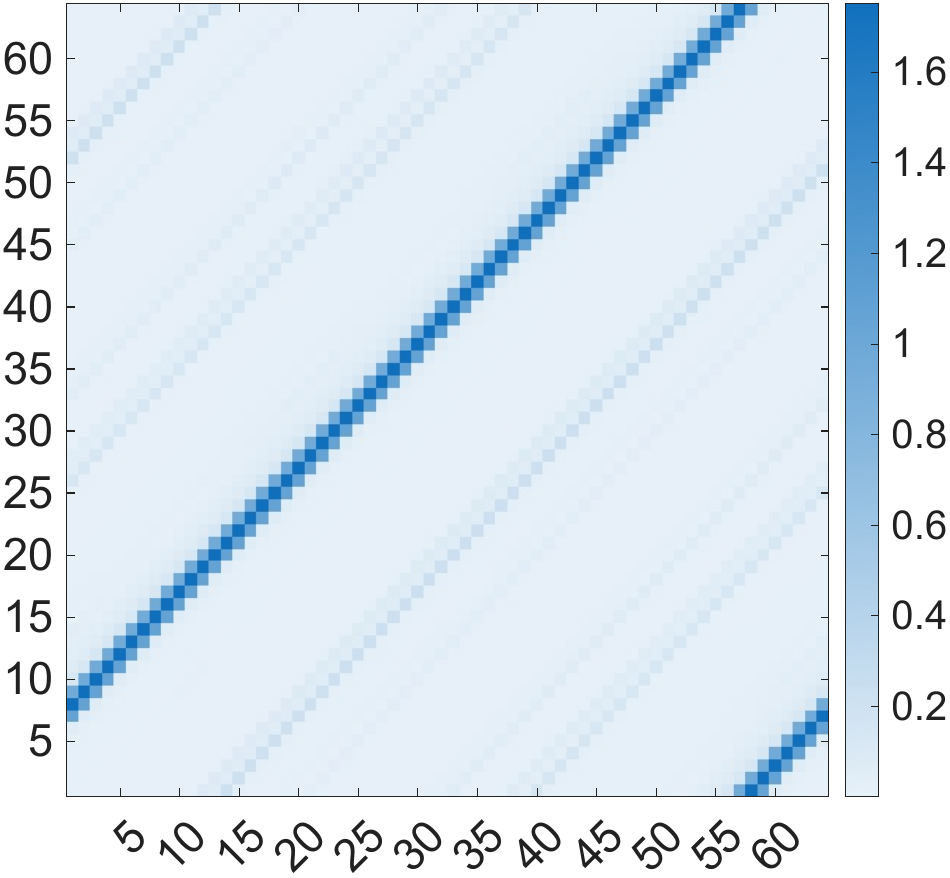}
\end{minipage}
}%
\centering
\caption{\LL{Comparison of time-domain $\bf{H}_t$ and the equivalent channels in MIMO-OFDM/OTFS/AFDM (Parameters: $J=U=1$, $N=64$ for OFDM and AFDM,  $(K=8, L=8)$ for OTFS, the number of multipath $P_{uj}=5$, and the velocity of device is $300$km/h).}}\label{Fig:H_eff}
\end{figure*}
After multicarrier demodulation, the received signal at the $u$-th antenna in \eqref{Eqn:Rev_y} is obtained as
\BE \label{Eqn:TF_y}
\bf{\bar{y}}_{u,t}= \bf{A}\bf{y}_{u,t}= \bf{H}_{{\rm{eff}},ut}\bf{s}_t+\bar{\bf{w}}_{u,t},
\EE
where effective channel $\bf{H}_{{\rm{eff}},ut}=\sum_{j=1}^{J}\bf{A}\bf{H}_{uj, t}\bf{A}^{\rm{H}}$, $\bf{s}_t=[\bf{s}_{1,t}^{\rm{T}}, ..., \bf{s}_{J,t}^{\rm{T}}]^{\rm{T}}$, and $\bar{\bf{w}}_{u,t}=\bf{A}\bf{w}_{u,t}$. \LL{Here $\bf{H}_{{\rm{eff}},ut}$ represents the effective channel corresponding to the superposition at the $u$-th receive antenna of the transmitted signals $\bf{s}_t$ from all $J$ transmit antennas.} {Note that the received signal expression in \eqref{Eqn:TF_y} is generic. Consequently, for the specific OFDM, OTFS, and AFDM systems, the demodulated signals are obtained by substituting $\bf{A}$ with their specific demodulation matrices.}

Furthermore, based on \eqref{Eqn:Rev_y}, the input-output model of the MIMO-multicarrier system in time domain is given by
\BE \label{Eqn:rev_mimo}
\bf{y}_t=\bf{H}_t\bf{x}_t+\bf{w}_t,
\EE
where
$\bf{y}_t=[\bf{y}_{1,t}^{\rm{T}}, ..., \bf{y}_{U,t}^{\rm{T}}]^{\rm{T}} \in \mathbb{C}^{UN\times 1}$, $\bf{H}_t=[\bf{H}_{1,t}^{\rm{T}}, ..., $
 $ \bf{H}_{U,t}^{\rm{T}}]^{\rm{T}}\in \mathbb{C}^{UN\times JN}$, and $\bf{w}_t=[\bf{w}_{1,t}^{\rm{T}}, ..., \bf{w}_{U,t}^{\rm{T}}]^{\rm{T}} \in \mathbb{C}^{UN \times 1}$.

Note that specific multicarrier modulation and demodulation schemes can be implemented (e.g., OFDM, OTFS, and AFDM) based on \eqref{Eqn:mcsignal} and \eqref{Eqn:TF_y}. Then,  effective signal detection and decoding techniques are employed to recover the message bits from \eqref{Eqn:TF_y}. {Here, we assume that the channel matrix $\bf{H}_t$ is perfectly known at the receiver but unknown at the transmitter.}


\subsubsection{OFDM}
\LL{The OFDM modulation matrix is the $N$-point inverse fast Fourier transform (IFFT) matrix $\bf{F}_{N}$, i.e., $\bf{A}=\bf{F}_{N}$. Based on \eqref{Eqn:mcsignal}, the time-domain signal is $\bf{x}^{\text{OFDM}}_{j,t}=\bf{F}_N^{\rm{H}}\bf{s}_{j,t}$. Correspondingly, based on \eqref{Eqn:TF_y}, the effective channel is $\bf{H}_{{\rm{eff}},ut}^{\text{OFDM}}= \sum_{j=1}^J\bf{F}_N \bf{H}_{uj,t} \bf{F}_N^{\text{H}}$. 
Due to the presence Doppler shifts, {the  diagonals of $\bf{F}_N \bf{H}_{uj,t} \bf{F}_N^{\text{H}}$ would be spread into a decaying band}, causing severe inter-carrier interference for OFDM symbol.}

\subsubsection{OTFS} 
\LL{Considering that OTFS is implemented via the inverse symplectic finite Fourier transform (ISFFT) and Heisenberg transform with a rectangular transmit pulse, the modulation matrix is expressed as
$\bf{A}=\bf{F}_L \otimes \bf{I}_{K}$ and the time-domain signal is $\bf{x}^{\text{OTFS}}_{j,t}=(\bf{F}_L^{\rm{H}} \otimes \bf{I}_{K}) \bf{s}_{j,t}$, where $N=KL$. Then, the corresponding effective channel is $\bf{H}_{{\rm{eff}},ut}^{\text{OTFS}}=\sum_{j=1}^{J}(\bf{F}_{L}\otimes \bf{I}_{K})\bf{H}_{uj,t}(\bf{F}_{L}^{\mr{H}}\otimes \bf{I}_{K})$.
}

\subsubsection{AFDM}
\LL{
The AFDM modulation matrix is the inverse discrete affine Fourier transform (IDAFT) matrix, i.e., $\bf{A}= \bf{\Lambda}_{c_2} \bf{F}_N \bf{\Lambda}_{c_1}$, thus the time-domain signal is $\bf{x}_{j,t}^{\text{AFDM}} =(\bf{\Lambda}_{c_1}^{\rm{H}} \bf{F}_N^{\rm{H}} \bf{\Lambda}_{c_2}^{\rm{H}}) \bf{s}_{j,t}$. The corresponding effective channel is $\bf{H}_{{\rm{eff}},ut}^{\text{AFDM}}=\sum_{j=1}^{J} \bf{\Lambda}_{c_2} \bf{F}_N \bf{\Lambda}_{c_1}\bf{H}_{uj,t} \bf{\Lambda}_{c_1}^{\rm{H}} \bf{F}_N^{\rm{H}} \bf{\Lambda}_{c_2}^{\rm{H}}$. Note that $c_1$ and $c_2$ are chosen to guarantee the full diversity of AFDM. 
}

\begin{table}[b] \scriptsize 
\caption{\LL{Comparisons of OFDM, OTFS, and AFDM in uncoded high-mobility systems.}}\label{Tab:Modu}
\centering
\begin{adjustbox}{width=\columnwidth}
\LL{\begin{tabular}{|c|c|c|c|c|c|c|}
\hline
\multirow{2}{*} {Modulation} &\multirow{2}{*} {\tabincell{c}{Transform\vspace{-0.15cm}\\Domain}} &\multirow{2}{*}{\tabincell{c}{Modulation\vspace{-0.15cm}\\Complexity}} & \multirow{2}{*}{\tabincell{c}{Equivalent  \vspace{-0.1cm}\\Channel}} & \multicolumn{2}{c|}{Robustness} & \multirow{2}{*} {\tabincell{c}{Diversity\vspace{-0.15cm}\\Type} }\\
\cline{5-6}
  & & & & Delay & Doppler &\\
\hline
\multirow{1}{*}{OFDM} & \multirow{1}{*}{Frequency} & \multirow{1}{*}{$\mathcal{O}(N\mr{log}N)$} & Dense & High & Low & None\\ 
\hline
OTFS & Delay-Doppler & {${\mathcal{O}(N\mr{log}L)}$} & Sparse & High & High& Time–Frequency\\ 
\hline
AFDM& DAFT & {$\mathcal{O}(N\mr{log}N)$} & Sparse  &  High & High& Time–Frequency\\  
\hline
\end{tabular}}
\end{adjustbox}
\end{table}

\subsubsection{Comparisons of OFDM, OTFS, and AFDM}
Table \ref{Tab:Modu} provides the comparisons of OFDM, OTFS, and AFDM in uncoded high-mobility systems. It is shown that both OFDM and AFDM are based on 1D transform with lower complexity, while OTFS is based on 2D transform. \LL{Fig.~\ref{Fig:H_eff} illustrates the super-sparsity of the time-domain channels through a comparative example between the original time-domain $\bf{H}_t$, and effective channels in symbol domain $\bf{H}_{{\rm{eff}},t}^{\text{OFDM}}$, $\bf{H}_{{\rm{eff}},t}^{\text{OTFS}}$, and $\bf{H}_{{\rm{eff}},t}^{\text{AFDM}}$, where $J=U=1$ and $N=64$ for OFDM and AFDM,  $(K=8, L=8)$ for OTFS, $P_{uj}=5$, and the velocity of device is $300$km/h.} It is noteworthy that the time-domain channel $\bf{H}_t$ is the sparsest, $\bf{H}_{{\rm{eff}}}^{\text{OFDM}}$ is dense and diagonal dominant, while $\bf{H}_{{\rm{eff}}}^{\text{OTFS}}$ and $\bf{H}_{{\rm{eff}}}^{\text{AFDM}}$ can distinct the different channel delays and Doppler shifts, and are regarded as sparse matrices by disregarding the smaller amplitude values.  Meanwhile, \LL{although uncoded OFDM is robust to channel delay, it is difficult to cope with the Doppler effect, thereby failing to achieve effective diversity in high mobility systems. In contrast, for OTFS and AFDM, the channel delay and Doppler can be distinguished, resulting in high diversity gain with excellent performance.}


\subsection{Key Challenges}
Most existing comparisons of OFDM, OTFS, and AFDM in communications have focused on uncoded systems, which still lack comprehensive investigations in the following aspects.
\begin{itemize}
\item How to develop low-complexity receivers for coded MIMO-multicarrier systems, especially large-scale systems, is still an open issue. For example, with OTFS, existing state-of-the-art CD-OAMP \cite{OTFS-OAMP} and DD-OAMP~\cite{DD-OAMP} detectors are still limited to the high-complexity LMMSE, while DD-MAMP \cite{MAMPOTFSconf} exploits the sparse DD domain channels but ignores the more sparse time-domain channels.
\item How to analyze the information-theoretical limit and develop the optimal code for MIMO-multicarrier systems is another challenge. Most existing works on OTFS and AFDM lack information-theoretical limit analysis.
 Meanwhile, the superiority of channel coding is frequently overlooked in existing receiver designs.
\end{itemize}

\section{Multi-Slot Cross-Domain MAMP Receiver and State Evolution}
In this section, we develop a multi-slot cross-domain MAMP (MS-CD-MAMP) receiver for coded MIMO-multicarrier systems and present the comparison of its complexity with existing state-of-the-art receivers. Then, the corresponding state evolution is presented.


\begin{figure}[t]\vspace{-0.2cm}
\centering  
\subfigure[MS-CD-MAMP receiver.]{
\includegraphics[width=0.9\columnwidth]{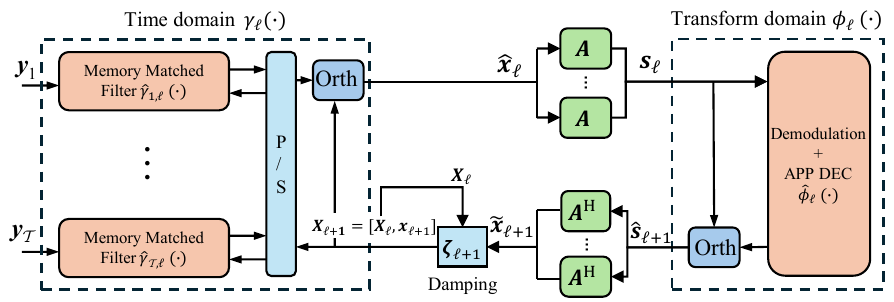}\label{Fig:MAMP0}}
\centering 
\vspace{-0.1cm}
\subfigure[State evolution.]{ 
\includegraphics[width=0.88\columnwidth]{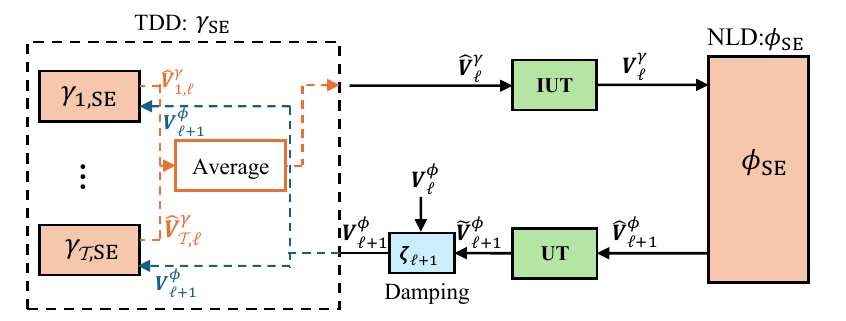}\label{Fig:MAMPSE}
}
\caption{Illustration of MS-CD-MAMP receiver and its state evolution.} \label{Fig:MAMP}
\end{figure}

\subsection{Multi-Slot Cross-Domain MAMP Receiver}
Considering the rapid change of doubly selective channels, transmission signals generally span multiple time slots while undergoing multiple channel fading. Since the AMP detector is restricted to IID channel matrices and may diverge for correlated channels, the OAMP detector was proposed to enhance the system performance in \cite{DD-OAMP,OTFS-OAMP,OAMP/VAMP}. However, its reliance on matrix inversion limits the exploitation of channel sparsity, leading to prohibitively high complexity in large-scale systems. 
\LL{To overcome these challenges, we develop a low-complexity MS-CD-MAMP receiver that fully exploits time-domain sparsity via a memory-matched filter, while achieving high reliability through joint channel coding across multiple time slots.}

\LL{As shown in Fig.~\ref{Fig:MAMP0}, the MS-CD-MAMP receiver consists of a time-domain detector $\gamma_{\ell}(\cdot)$, an inverse unitary transform, a nonlinear detector $\phi_{\ell}(\cdot)$ in the multicarrier symbol domain, a unitary transform, and a damping operation, where ${\ell}$ is the iteration index. In particular, $\gamma_{\ell}(\cdot)$ is composed of $\mathcal{T}$ memory matched filters $\hat{\gamma}_{t,\ell}(\cdot)$, parallel-to-serial (P/S) conversion, and orthogonalization, with $\mathcal{T}$ denoting the total number of time slots occupied by the transmitted signals. Notably, by leveraging the received signals of each time slot and the sparsity of the time-domain channel matrices $\{\bf{H}_t\}$, all per-slot local estimations are performed within $\{\hat{\gamma}_{t,\ell}(\cdot)\}$. Meanwhile, the estimated input and output signals across the $\mathcal{T}$ time slots are jointly orthogonalized after P/S conversion, thereby eliminating inter-slot asymmetry and the input–output estimation errors correlation during the iteration.} \LL{$\phi_{\ell}(\cdot)$ consists of a demodulator $\hat{\phi}_{\ell}(\cdot)$ and an orthogonalization operation. Detailed descriptions are given below.}

\subsubsection{Time-Domain Detection (TDD)}
\LL{For received signals $\bf{y}_1, ..., \bf{y}_{\mathcal{T}}$, estimated signal $\hat{\bf{x}}_{\ell}\in \mathbb{C}^{M\times 1}$ at the ${\ell}$-th iteration is obtained through TDD $\gamma_{\ell}(\cdot)$, using  \emph{a priori} signal $\bf{X}_{\ell}=[\bf{x}_1, ..., \bf{x}_{\ell}]\in\mathbb{C}^{M\times \ell}$ derived from all previous iterations, where $M=NJ\mathcal{T}$, $\bf{x}_{\ell}=[(\bf{x}_{\ell}^{1})^{\rm{T}}, ..., (\bf{x}_{\ell}^{\mathcal{T}})^{\rm{T}}]^{\rm{T}}\in \mathbb{C}^{M\times 1}$, and $\bf{x}_{\ell}^{t}\in \mathbb{C}^{NJ\times 1}$ denotes the \emph{a priori} estimations for the $t$-th slot at the $\ell$-th iteration. Starting with ${\ell}=1$ and $\bf{X}_1=\bf{0}$, 
\BE\label{Eqn:MLD}
\bf{\hat{x}}_{\ell}=\gamma_{\ell}(\bf{X}_{\ell})=\frac{1}{\varepsilon_{\ell}^\gamma}([\hat{\gamma}_{1,\ell}({\bf{X}}_{\ell}^1), ..., \hat{\gamma}_{\mathcal{T},\ell}({\bf{X}}_{\ell}^{\mathcal{T}})]-\bf{X}_{\ell} \bf{p}_{\ell}),
\EE
where $\bf{X}_{\ell}^t=[\bf{x}_{1}^t, ..., \bf{x}_{\ell}^t] \in\mathbb{C}^{NJ\times \ell}$ denotes the estimated signals of the $t$-th time slot from all previous iterations, $\varepsilon_{\ell}^\gamma$ and $\bf{p}_{\ell}$ are the normalized and orthogonal parameters respectively, and $\hat{\gamma}_{t, \ell}(\cdot)$ is given by \cite[Equation (37)]{LeiMAMP}
}
\BS\label{Eqn:mmf}
\BE \label{Eqn:mmf0}
\hat{\gamma}_{t, \ell}(\bf{X}_{\ell}^t)=\bf{H}_t^{\mathrm{H}} \tilde{\gamma}_{t,\ell}(\bf{X}_{\ell}^t) 
\EE
with 
\BE\label{Eqn:mmf1}
\tilde{\gamma}_{t, \ell}(\bf{X}_{\ell}^t)=\theta_{\ell}^t \bf{B}_t \tilde{\gamma}_{t,{\ell}-1}(\bf{X}_{{\ell}-1}^t)+\xi^t_{\ell}(\bf{y}_t-\bf{H}_t \bf{x}_{\ell}^t),
\EE
\ES
where $\tilde{\gamma}_{t,0}(\bf{X}_0^t)=\bf{0}, \bf{B}_t=\lambda^{\dagger}_t \bf{I}-\bf{H}_t \bf{H}_t^{\mathrm{H}}$ with $\lambda^{\dagger}_t=(\lambda^{\min}_t+\lambda^{\max }_t)/2$, $\lambda^{\min}_t$ and $\lambda^{\max}_t$ denote the minimal and maximal eigenvalues of $\bf{H}_t \bf{H}_t^{\mathrm{H}}$, respectively\footnote{{If they are unavailable, we can adopt a low-complexity approximation of $\lambda_{\text{min}}$ and $\lambda_{\text{max}}$ provided in \cite[Equation (7)]{LeiMAMP}, in which this approximation method has no effect on the complexity order of MAMP and the MAMP with approximate eigenvalues can achieve the same performances as that with exact eigenvalue \cite{LeiMAMP,CodeMAMP2023}.}}. 
\LL{Similar to \cite{LeiMAMP}, we employ an iterative multi-slot memory-matched filter to approximate the LMMSE estimates as in \eqref{Eqn:lmmse_app}.} Relaxation parameter $\left\{\theta_{\ell}^{t}\right\}$  is set to promote the convergence of the MAMP receiver. $\{\xi_{\ell}^t\}$ is optimized to accelerate the convergence.  The detailed parameter calculations are given as follows.
\begin{itemize}
    \item $\{\theta^t_{\ell}\}$: $\theta^t_{\ell}=(\lambda_{t}^{\dagger}+\rho_{\ell})^{-1},$
    where $\rho_{\ell}=\sigma^2 / v_{\ell, \ell}^{{\phi}}$, $\sigma^2$ is the noise variance,  $v_{\ell, \ell}^{{\phi}}=\frac{1}{M} \mathrm{E}\{\|\bf{x}_{\ell}-\bf{x}\|^2\}$, and $\bf{x}=\{\bf{x}_{j,t}\}$ denotes the all time-domain transmitted signals in \eqref{Eqn:mcsignal}, $j=1, ..., J$, $t=1, ..., \mathcal{T}$.
    \item $\{\xi^t_{\ell}\}$: $\{\xi^t_{\ell}\}$ is optimized jointly over all time slots to minimize the variances of $\{\bf{\hat{x}}^t_{\ell}\}$ similar to \cite[Section V-B]{LeiMAMP}. 
    \item $\varepsilon_{\ell}^\gamma$ and $\bf{p}_{\ell}$: 
{Following \cite{LeiMAMP}, for $t\ge 0$, we define
\BS\label{Eqn:sym_defs}\begin{align} 
 \bf{W}^{\ell}_t &\equiv  \bf{H}^{\rm H}_t\bf{B}^{\ell}_t\bf{H}_t,\label{Eqn:sym_defs_w}\\
 b^{\ell}_t &\equiv \tfrac{1}{JN}{\rm tr}\{\bf{B}_t^{\ell}\} = \textstyle\sum_{i=0}^{\ell} \binom{\ell}{i} (-1)^i(\lambda_t^\dag)^{\ell-i}\lambda_t^{\ell}, \label{Eqn:sys_defs_b}\\  
   w_t^{\ell} &\equiv  \tfrac{1}{JN}{\rm tr}\{\bf{W}_t^{\ell}\}= \lambda^\dag_t b_{t}^{\ell}- b_{t+1}^{\ell},\label{Eqn:b_w}
\end{align}  \ES 
For $1\le i\le \ell$,   
\BS\label{Eqn:orth_parameters} \begin{align}
   \vartheta_{\ell, i}^{t}  &\equiv \left\{\!\!\! \begin{array}{ll}
    {\xi}^{t}_{\ell}, & i=\ell   \vspace{0.15cm}\\
     {\xi}_i^{t}  \textstyle\prod_{\tau=i+1}^{\ell}\theta_\tau^t, & i<\ell
  \end{array}\right.,\end{align}\begin{align}
  p_{\ell, i}^{t}&\equiv  -\vartheta_{\ell, i}^{t}   w_{t-i}^{\ell},\label{Eqn:orth_parametersb} \\ 
  {\varepsilon}^\gamma_t &\equiv  -\textstyle\sum_{i=1}^{\ell}  p_{t, i}^{\ell}.  \label{Eqn:orth_parametersc}
\end{align} \ES     
 Furthermore, $\vartheta_{\ell, i}^t=1$ if $i>\ell$. }
 \end{itemize}

Therefore, the output average variance of $\gamma_{\ell}(\cdot)$ is 
\BE\label{Eqn:TDvar}
\hat{v}_{\ell, \ell}^\gamma = \frac{1}{M}\mathrm{E}\{\|\gamma_{\ell}(\bf{X}_\ell)-\bf{x}\|^2\}.
\EE

\subsubsection{Inverse Unitary Transform (IUT)}
Through the IUT, signal $\bf{s}_{\ell}$ and variance $v_{\ell, \ell}^\gamma$ are obtained as
{
\BE\label{Eqn:demo}
\begin{aligned}
\bf{s}_{\ell} & =(\bf{I}_{J\mathcal{T}} \otimes \bf{A})\bf{\hat{x}}_{\ell}, \\
\bf{V}_{\ell, \ell}^\gamma & =\hat{v}_{\ell, \ell}^\gamma{(\bf{I}_{J\mathcal{T}} \otimes \bf{A})}^{\rm{H}}(\bf{I}_{J\mathcal{T}} \otimes \bf{A})=\hat{v}_{\ell, \ell}^\gamma\bf{I}_{JT}.
\end{aligned}
\EE
As a result, the variance of each element in $\bf{s}_{\ell}$ is  $v_{\ell, \ell}^{\gamma}=\hat{v}_{\ell, \ell}^{\gamma}$.
}

\subsubsection{Non-Linear Detection (NLD)} 
The NLD $\phi_{\ell}(\cdot)$ composes of $\hat{\phi}_{\ell}(\cdot)$ and orthogonalization, where $\hat{\phi}_{\ell}(\cdot)$ includes a symbol-by-symbol MMSE demodulation and \emph{a-posteriori probability} (APP) decoding. The output estimation of $\phi_{\ell}(\cdot)$ is
\BE\label{Eqn:nld}
\bf{\hat{s}}_{\ell+1}=\phi_{\ell}(\bf{s}_{\ell})=\frac{1}{\varepsilon_{\ell}^\phi}(\hat{\phi}_{\ell}(\bf{s}_{\ell})-p_{\ell}^\phi \bf{s}_{\ell}),
\EE
where $\varepsilon_{\ell}^\phi$ is a constant determined by minimizing the mean squared error (MSE) of $\phi_{\ell}(\cdot)$, {$\hat{\phi}_{\ell} (\bf{s}_{\ell})=\mr{E}\{\bf{s}|\bf{s}_{\ell}\}$}, and $p_{\ell}^\phi={\mr E}\{\frac{\partial \hat{\phi}_{\ell}}{\partial \bf{s}_{\ell}}\}$ is the orthogonal parameter, respectively.

\LL{In the symbol-by-symbol MMSE demodulation, the APP of the $k$-th symbol of $\bf{s}_{\ell}$ is 
\BE\label{Eqn:mmse_demo}
{P}_{\rm APP}(\bf{s}_{\ell,k}=\bf{\chi}) \propto \exp \left\{-\frac{\|\bf{\chi}-\bf{s}_{\ell,k}\|_2^2}{v_{\ell, \ell}^{\gamma}}\right\},
\EE
where $k=1,..., M$, $\bf{\chi} \in \mathcal{S}$, and $\mathcal{S}$ denotes the constellation set. On this basis, after channel decoding, the \emph{a posteriori} estimation of $\hat{\phi}_{\ell}(\cdot)$ can be obtained by using the output APP, i.e., $\hat{\phi}_{\ell} (\bf{s}_{\ell})=\mr{E}\{\bf{s}|\bf{s}_{\ell}\}$.
}




The output average variance of $\phi_{\ell}(\cdot)$ is
\BE\label{Eqn:nldvar}
\hat{v}_{\ell+1, \ell+1}^\phi \equiv \frac{1}{M} \mathrm{E}\{\|\phi_{\ell}(\bf{s}_{\ell})-\bf{s}\|^2\},
\EE
where $\bf{s}=\{\bf{s}_{j,t}\}$ denotes the all transmitted signals in \eqref{Eqn:mcsignal}, $j=1, ..., J$, $t=1, ..., \mathcal{T}$.

\subsubsection{Unitary Transform (UT)} Through UT,  signal $\bf{\tilde{x}}_{\ell+1}$ and variance $\tilde{v}_{\ell+1, \ell+1}^\phi$ are obtained as:
{
\BE\label{Eqn:mt_mv}
\begin{aligned}
\bf{\tilde{x}}_{\ell+1} & =(\bf{I}_{J\mathcal{T}} \otimes \bf{A}^{\rm H})\bf{\hat{s}}_{\ell+1}, \\
\bf{\tilde{V}}_{\ell+1, \ell+1}^\phi & \!=\!\hat{v}_{\ell+1, \ell+1}^\phi(\bf{I}_{J\mathcal{T}}\!\!\otimes\!\!\bf{A}^{\rm{H}})(\bf{I}_{J\mathcal{T}} \!\!\otimes \!\!\bf{A}^{\rm{H}})^{\rm{H}}\!=\!{\hat{v}}_{\ell+1, \ell+1}^\phi\bf{I}_{JT},
\end{aligned}
\EE
As a result, the variance of each element of $\bf{\tilde{x}}_{\ell+1}$ is  $\tilde{v}_{\ell+1, \ell+1}^\phi={\hat{v}}_{\ell+1, \ell+1}^\phi$, which are fed back to the time-domain detector $\gamma_{\ell}(\cdot)$.
}
\LL{
\subsubsection{Damping}
To guarantee and improve the convergence of MS-CD-MAMP in principle, a damping strategy is applied to obtain $\bf{x}_{\ell+1}$ by all estimated
signals $\bf{X}_{\ell}$ up to the $\ell$-th iteration and the latest estimated signal $\tilde{\bf{x}}_{\ell+1}$ after UT, i.e.,
\BE\label{Eqn:damp}
\bf{x}_{\ell+1} = [\bf{X}_{\ell}, \tilde{\bf{x}}_{\ell+1}] \cdot \bf{\zeta}_{\ell+1},
\EE
where $\bf{\zeta}_{\ell+1}$ is a damping vector for the linear weighted superposition of all estimations. Meanwhile, $\bf{x}_{\ell+1}$ is stacked into $\bf{X}_{\ell+1}=[\bf{X}_{\ell},\bf{x}_{\ell+1}]$ before entering TDD.
}

\LL{Similar to \cite{LeiMAMP}, ``full orthogonality" in \eqref{Eqn:MLD} and \eqref{Eqn:nld} eliminates the correlation among input–output errors of the local estimator across iterations and ensures that the input to the NLD exhibits asymptotic IID Gaussianity in Lemma\ref{Lem:iid}, thereby enabling the design of a locally MMSE-optimal NLD. More importantly, by leveraging orthogonality, state evolution can accurately characterize the asymptotic performance of the MS-CD-MAMP receiver, facilitating the analysis of its maximum achievable rate and guiding optimal coding design.}


\begin{assumption}[Uniformly Lipschitz-Continuous APP Decoder for forward error correction (FEC) codes]\label{Asp:fec}
The APP decoder of FEC code is assumed to be uniformly Lipschitz-continuous.
\end{assumption}

\begin{lemma}[Asymptotically IID Gaussianity]\label{Lem:iid}
Assume that ${\gamma_\ell(\cdot)} $ and ${\phi_\ell(\cdot)}$ are uniformly Lipschitz-continuous estimators.
Let $\bf{X}=\bf{x} \cdot \bf{1}^T$ with an all-ones vector $\bf{1}$ of proper size and $\bf{S}=\bf{s} \cdot \bf{1}^T$. The ``full orthogonality" guarantees the asymptotically IID Gaussianity of estimate errors, i.e., 
\BE
{\bf{X}_{\ell}}=\bf{X}+\bf{Z}_{\ell}^\phi,\quad
{\bf{S}_{\ell}}=\bf{S}+\bf{Z}_{\ell}^\gamma,
\EE
where $\bf{Z}_{\ell}^\gamma=[\bf{z}_1^\gamma, \ldots, \bf{z}_{\ell}^\gamma]$ and $\bf{Z}_{\ell}^\phi=[\bf{z}_1^\phi, \ldots, \bf{z}_{\ell}^\phi]$ are column-wise IID Gaussianity and row-wise joint-Gaussian and independent of $\bf{x}$, i.e., 
$\bf{z}_{\ell}^\gamma \sim  \mathcal{CN}(\mathbf{0}, v_{{\ell}, {\ell}}^\gamma \bf{I})$ with $\mathrm{E}\{\bf{z}_{\ell}^\gamma(\bf{z}_{{\ell}^{\prime}}^\gamma)^{\mathrm{H}}\}=v_{{\ell}, {\ell}^{\prime}}^\gamma \bf{I}$ and $\bf{z}_{\ell}^\phi \sim \mathcal{C N}(\mathbf{0}, v_{{\ell}, {\ell}}^\phi \bf{I})$ with $\mathrm{E}\{\bf{z}_{\ell}^\phi(\bf{z}_{{\ell}^{\prime}}^\phi)^{\mathrm{H}}\}=v_{{\ell}, {\ell}^{\prime}}^\phi \bf{I}$.
\end{lemma}

\vspace{-0.2cm}
\subsection{Complexity Comparison}
Table \ref{Tab:complexity} presents the complexity comparisons between the proposed MS-CD-MAMP receiver and existing state-of-the-art receivers in MIMO-multicarrier systems, including GMP~\cite{GMP2018}\cite{GeYao2021}, CD-OAMP\cite{OTFS-OAMP}, DD-OAMP~\cite{DD-OAMP}\cite{OAMP/VAMP}, DD-MAMP\cite{MAMPOTFSconf}, LMMSE\cite{LMMSE&ZF}\cite{AfdmLow}, where $\mathcal{L}$ denotes the maximum iteration number, {maximum channel taps $\mathcal{I}\ll N $}, and LDPC codes are employed (See details in the section \ref{sec:NR}). 
Define $S_{\bf{H}_{{\rm{eff}},t}}$ and $S_{\bf{B}_{\rm{eff}, t}}$ as the average number of non-zero entries in each row of symbol-domain equivalent channel  $\bf{H}_{\rm{eff},t}=[\bf{H}_{{\rm{eff}},1t}^{\rm{T}}, ..., \bf{H}_{{\rm{eff}},Ut}^{\rm{T}}]$ in \eqref{Eqn:TF_y} and $\bf{B}_{\rm{eff}, t}$, where $\bf{B}_{\rm{eff}, t}$ is obtained based on $\bf{H}_{\rm{eff},t}$ similar to \eqref{Eqn:mmf1}.
Note that the complexity of the LDPC decoder is approximated as $\mathcal{O}(M\mathcal{L})$ using the sum-product decoding algorithm~\cite{ryan2009channel}. As a result, the complexity of different receivers is {dominated} by the detector.\LL{ The time complexity of the LMMSE receiver is limited by the matrix inversion, which is $\mathcal{O}((NJ)^3 \mathcal{T})$, and correspondingly, the space complexity is $\mathcal{O}\bigl(N^{2}UJ\mathcal{T}\bigr)$. Similarly, the time complexity of CD-OAMP and DD-OAMP is respectively $\mathcal{O}((NJ)^3\mathcal{L}\mathcal{T}+2NJ\mathcal{L}\log{NJ}+M\mathcal{L})$ and $\mathcal{O}((NJ)^3\mathcal{L}\mathcal{T}+M\mathcal{L})$, which is also limited by matrix inversion of iterative LMMSE and unitary transform in CD-OAMP. Meanwhile, the dominant term in space complexity of both CD-OAMP and DD-OAMP is also attributable to LMMSE. The complexity of GMP and DD-MAMP is limited by the sparsity of DD domain channel matrix, which is $\mathcal{O}((NUS_{\bf{H}_{{\rm{eff}},t}}\mathcal{L}\mathcal{T}+M\mathcal{L})$ and $\mathcal{O}(NU(S_{\bf{B}_{\rm{eff}, t}}+3S_{\bf{H}_{{\rm{eff}},t}}+1)\mathcal{L}\mathcal{T}+M\mathcal{L})$ respectively. The space complexity of DD-MAMP requires consideration of the impact of memory filters, which is $\mathcal{O}\bigl(N^{2}UJ+M\mathcal{L}+\mathcal{L}^{2}\mathcal{T}\bigr)$.  In contrast, the proposed MS-CD-MAMP receiver can fully exploit the {sparsity of} time-domain channel matrix, whose time complexity is as low as $\mathcal{O}(\mathcal{I}NJ\mathcal{L}\mathcal{T}+2NJ\mathcal{L}\log{NJ}+M\mathcal{L})$, and the space complexity is correspondingly reduced to $\mathcal{O}\bigl((\mathcal{I}NUJ+M\mathcal{L}+\mathcal{L}^{2})\mathcal{T}\bigr)$. }

\LL{
\begin{table}[htbp]
\tiny
\centering
\caption{Complexity Comparison of Different Receivers.}
\label{Tab:complexity}
\LL{
\begin{tabularx}{\linewidth}{|c|>{\centering\arraybackslash}X|>{\centering\arraybackslash}X|}
\hline
Algorithm & Time complexity & Space complexity \\ \hline
LMMSE \cite{LMMSE&ZF,AfdmLow} &
$\mathcal{O}\bigl((NJ)^3 \mathcal{T} + M\mathcal{L}\bigr)$ &
$\mathcal{O}\bigl(N^{2}UJ\mathcal{T}\bigr)$ \\ \hline
GMP \cite{GMP2018,GeYao2021} &
$\mathcal{O}\bigl(NUS_{\mathbf{H}_{\text{eff},t}}\mathcal{L}\mathcal{T}+M\mathcal{L}\bigr)$ &
$\mathcal{O}\bigl(N^{2}UJ\mathcal{T}\bigr)$ \\ \hline
CD-OAMP \cite{OTFS-OAMP} &
$\mathcal{O}\bigl((NJ)^3\mathcal{L}\mathcal{T}+2NJ\mathcal{L}\log NJ+M\mathcal{L}\bigr)$ &
$\mathcal{O}\bigl(N^{2}UJ\mathcal{T}\bigr)$ \\ \hline
DD-OAMP \cite{DD-OAMP,OAMP/VAMP} &
$\mathcal{O}\bigl(((NJ)^3\mathcal{L}\mathcal{T}+M\mathcal{L})\mathcal{T}\bigr)$ &
$\mathcal{O}\bigl(N^{2}UJ\mathcal{T}\bigr)$ \\ \hline
DD-MAMP \cite{MAMPOTFSconf} &
$\mathcal{O}\bigl(NU(S_{\mathbf{B}_{\text{eff},t}}+3S_{\mathbf{H}_{\text{eff},t}}+1)\mathcal{L}\mathcal{T}+M\mathcal{L}\bigr)$ &
$\mathcal{O}\bigl((N^{2}UJ+M\mathcal{L}+\mathcal{L}^{2})\mathcal{T}\bigr)$ \\ \hline
\makecell{MS-CD-MAMP\\(Proposed)} &
$\mathcal{O}\bigl(\mathcal{I}NJ\mathcal{L}\mathcal{T}+2NJ\mathcal{L}\log NJ+M\mathcal{L}\bigr)$ &
$\mathcal{O}\bigl((\mathcal{I}NUJ+M\mathcal{L}+\mathcal{L}^{2})\mathcal{T}\bigr)$ \\ \hline
\end{tabularx}
}
\end{table}}

\vspace{-0.7cm}
\subsection{State Evolution of MS-CD-MAMP Receiver}
Since the multiple memory matched filters $\{\hat{\gamma}_{t, \ell}(\cdot)\}$ are employed in \eqref{Eqn:MLD}, the asymptotic performance can be evaluated by using SE based on the covariance matrix. Based on the IID Gaussianity in Lemma~\ref{Lem:iid}, as shown in Fig. \ref{Fig:MAMPSE}, the asymptotic MSE performance of MS-CD-MAMP can be predicted by the MSE functions $\gamma_{\mathrm{SE}}(\cdot)$ and $\phi^{\mathcal{C}}_{\mathrm{SE}}(\cdot)$, i.e., 
\BS\label{Eqn:MAMPSE}
\begin{align}  \label{Eqn:se_tdd}
\!\!\!\text{TDD:} &\; {\bf{\hat{V}}_{\ell}^\gamma =\gamma_{\mathrm{SE}}(\tilde{\bf{V}}_{\ell}^{\phi}) =\frac{1}{\mathcal{T}}\sum_{t=1}^{\mathcal{T}}\gamma_{t,\mathrm{SE}}(f_{\mathrm {d}}(\tilde{\bf{V}}_{\ell}^{\phi}, {\bf{V}}_{\ell-1}^{\phi}))}\\ \label{Eqn:se_md}
\text{IUT:} &\; \bf{V}_{\ell}^{\gamma}=(\bf{I}_{J\mathcal{T}} \otimes \bf{A})\cdot\bf{\hat{V}}_{\ell}^{\gamma}\cdot(\bf{I}_{J\mathcal{T}} \otimes \bf{A}^{\rm{H}})\\
\label{Eqn:se_nld}
\text{NLD:}& \; \bf{\hat{V}}_{{\ell}+1}^{\phi} =\phi_{\mathrm{SE}}^{\mathcal{C}}(\bf{V}_{\ell}^{\gamma})\\
\label{Eqn:se_mt}
\text{UT:} & \; \tilde{\bf{V}}_{\ell+1}^{\phi}=(\bf{I}_{J\mathcal{T}} \otimes \bf{A}^{\rm{H}})\cdot\bf{\hat{V}}_{\ell+1}^{\phi} \cdot(\bf{I}_{J\mathcal{T}} \otimes \bf{A})
\end{align}  
\ES
where $\gamma_{\mathrm{SE}}(\cdot)$ is obtained by averaging over multiple slot-wise SE functions $\{\gamma_{t,\mathrm{SE}}(\cdot)\}$, $\{\hat{\bf{V}}_{\ell}^\gamma, {\bf{V}}_{\ell}^\phi\}$ and $\hat{\bf{V}}_{\ell}^{\phi}$ are the covariance matrices for $\bf{X}$ and $\bf{S}$ during the iterations, i.e., 
\BS
\begin{align}
    \hat{\bf{V}}_{\ell}^\gamma &=  \langle \hat{\bf{X}}_{\ell}-\bf{X} \mid \hat{\bf{X}}_{\ell}-\bf{X}\rangle, \\
    {\bf{V}}_{\ell}^{\phi} &=  \langle {\bf{X}}_{\ell}-\bf{X} \mid {\bf{X}}_{\ell}-\bf{X}\rangle,\\
    \hat{\bf{V}}_{\ell}^\phi &=  \langle \hat{\bf{S}}_{\ell}-\bf{S} \mid \hat{\bf{S}}_{\ell}-\bf{S} \rangle,
\end{align}
\ES
with $\hat{\bf{X}}_{\ell}=[\hat{\bf{x}}_{1}, ..., \hat{\bf{x}}_{\ell}]$ and $\hat{\bf{S}}_{\ell}=[\hat{\bf{s}}_1, ..., \hat{\bf{s}}_{\ell}]$.

In \eqref{Eqn:se_tdd}, $f_{\mathrm {d}}(\cdot)$ denotes the damping function based on  $\tilde{\bf{V}}_{\ell}^{\phi}$ and ${\bf{V}}_{\ell-1}^{\phi}$, i.e., ${\bf{V}}_{\ell}^{\phi}=f_{\rm d}(\tilde{\bf{V}}_{\ell}^{\phi}, {\bf{V}}_{\ell-1}^{\phi})$.
That is, the $\ell$-th row of  ${\bf{V}}_{\ell}^{\phi}$ is 
\BE
(\bf{v}_{{\ell}}^{\phi})^{\mathrm{T}}=\bf{\Lambda}_{{\ell}}^{\mathrm{H}} \bf{\mathcal{V}}_{{\ell}}[
\bf{I}_{{\ell} \times {(\ell-1)}}\;\;\bf{\Lambda}_{{\ell}}],
\EE
where $\bf{I}_{{\ell} \times {(\ell-1)}}$ is a ${\ell} \times {(\ell-1)}$ sub-matrix of $\bf{I}_{{\ell} \times {\ell}}$ excluding the last column, $\bf{\mathcal{V}}_{\ell}$ denotes the error covariance matrix of $\left\{\bf{x}_1, \ldots, \bf{x}_{\ell-1}, \tilde{\bf{x}}_{{\ell}}\right\}$, i.e., 
\BE\label{Eqn:V_phi_a}
 \bf{\mathcal{V}}_{\ell} \equiv \left[
\begin{array}{cc}
        \bf{V}_{\ell-1}^{{\phi}}   &  \begin{array}{c}
             \tilde{v}^{\phi}_{1,\ell}  \\
            \vdots  \end{array}  \\ 
        \begin{array}{cc}
             \tilde{v}^{\phi}_{\ell, 1} &    \cdots 
        \end{array}  &  \tilde{v}^{\phi}_{\ell,\ell}
\end{array}
\right]_{\ell \times \ell},
\EE
$\bf{v}_{{\ell}}^{\phi}=[v_{{\ell},1}^{\phi}, \ldots, v_{{\ell}, {\ell}}^{\phi}]^{\mathrm{T}}$, and $\tilde{v}_{\ell,\ell\prime}=\frac{1}{M}{\mr E}\{[\tilde{\bf{x}}_{\ell-1}-\bf{x}]^{\rm{H}}[\bf{x}_{\ell \prime}-\bf{x}]\}$
(see details in~\cite[Appendix C and Appendix H]{LeiMAMP}).

For \eqref{Eqn:se_md}, we can obtain 
\begin{align*}
    \bf{V}_{\ell}^{\gamma}&=\langle {\bf{S}}_{\ell}-\bf{S} \mid {\bf{S}}_{\ell}-\bf{S} \rangle, \\
    &=\frac{1}{M}(\bf{S}_\ell-\bf{S})^{\rm H}(\bf{S}_{\ell}-\bf{S})\\
    &=\frac{1}{M}\left[(\bf{I}_{J\mathcal{T}} \otimes \bf{A})\cdot \hat{\bf{X}}_\ell-(\bf{I}_{J\mathcal{T}} \otimes \bf{A}) \cdot {\bf{X}}\right]^{\rm H}\cdot\\&\left[(\bf{I}_{J\mathcal{T}} \otimes \bf{A})\cdot \hat{\bf{X}}_\ell-(\bf{I}_{J\mathcal{T}} \otimes \bf{A}) \cdot \bf{X}\right]\\
    &=\frac{1}{M}( \hat{\bf{X}}_\ell- \bf{X})^{\rm H}( \hat{\bf{X}}_\ell- \bf{X})=\hat{\bf{V}}_{\ell}^\gamma.
\end{align*}
Similarly, for \eqref{Eqn:se_mt}, we have $\tilde{\bf{V}}_{\ell+1}^{\phi}=\bf{\hat{V}}_{\ell+1}^{\phi}$. 

As a result, the MSE functions of MS-CD-MAMP are rewritten as
\BS\label{Eqn:MAMPSE2}
\begin{align}  
\!\!\!\text{TDD:}&\; \bf{V}_{\ell}^\gamma =\gamma_{\mathrm{SE}}(\tilde{\bf{V}}_{\ell}^{\phi}) =\frac{1}{\mathcal{T}}\sum_{t=1}^{\mathcal{T}}\gamma_{t,\mathrm{SE}}(f_{\mathrm {d}}(\tilde{\bf{V}}_{\ell}^{\phi}, {\bf{V}}_{\ell-1}^{\phi})),\\
\!\!\!\text{NLD:}&\; \tilde{\bf{V}}_{{\ell}+1}^{\phi} =\phi_{\mathrm{SE}}^{\mathcal{C}}(\bf{V}_{\ell}^{\gamma}).
\end{align}  
\ES

Based on the simplified SE in \eqref{Eqn:MAMPSE2}, the following theorem is given to show that the equivalent MSE performance of cross-domain MAMP and symbol-domain MAMP~\cite{MAMPOTFSconf}.

\begin{theorem}[MSE Equivalency of Cross-Domain and Symbol-Domain MAMP\cite{MAMPOTFSconf}]\label{The:MSE_eq}
For multicarrier modulations based on unitary matrix transforms, the SE of cross-domain MAMP and symbol-domain MAMP \cite{MAMPOTFSconf} is the same, i.e., they can achieve the same MSE performance.
\end{theorem}

It is worth noting that MS-CD-MAMP's memory matched filters rely on the estimations of all previous iterations to produce the output at the current iteration. If only the current estimation is considered and 
\BE\label{Eqn:lmmse_app}
 \tilde{\gamma}_{t,\ell}(\bf{X}_{\ell}^t) = \frac{\xi^t_{\ell}}{\theta^t_{\ell}} \left[\left(\frac{1}{\theta^t_{\ell}}-\lambda_t^{\dagger}\right) \bf{I}+\bf{H_t}\bf{H_t}^{\mathrm{H}}\right]^{-1}\left(\bf{y_t}-\bf{H_t} \bf{x}^t_{\ell}\right),
\EE
the MS-CD-MAMP is reduced to the non-memory MS-CD-OAMP/VAMP. As a result, based on Theorem~\ref{The:MSE_eq}, the following corollary is obtained straightforwardly.
\begin{corollary}[MSE Equivalency of Cross-Domain and Symbol-Domain OAMP/VAMP in \cite{OTFS-OAMP} and \cite{DD-OAMP}]\label{cor:mseOAMP}
For unitary matrix transform-based multicarrier modulations, the SE of the cross-domain and symbol-domain OAMP/VAMP is the same, thereby achieving the same MSE performance.
\end{corollary}

{\emph{Remark:} Since the subsequent achievable rate analysis and optimal coding principle for the MS-CD-MAMP rely on precise SE analysis, it is essential that the decoder satisfies the Lipschitz continuity condition. Note that the belief propagation decoder of LDPC code is shown to satisfy a stronger Lipschitz-continuous in \cite[Appendix B]{ebert2023sparse}  under a sub-girth condition (widely used in LDPC codes) that fewer message passing iterations are performed on the factor graph of the LDPC code than the shortest cycle of the same graph per decoding iteration. This implies that the SE based on LDPC decoding is accurate, which has been verified by simulation results in \cite{LeiTIT2021,MaTWC2019,LeiOptOAMP,YuhaoTcom2022}.  As a result, a kind of LDPC code is designed based on the SE in the numerical results of this paper. Although there is no rigorous proof for other types of FEC codes, we conjecture that the APP decoder should be uniformly Lipschitz-continuous for the majority of FEC codes (e.g., Turbo code, Polar code, etc.), which is an important topic for future work for the design of other types codes.}

\section{Achievable Rates Analysis and Coding Principle of MS-CD-MAMP Receiver} 
In this section, we first present the achievable rate analysis and the optimal coding principle for MS-CD-MAMP, followed by the maximum achievable rate comparisons of MS-CD-MAMP with various system parameters.


\subsection{Theoretical Achievable Rate Analysis and Optimal Coding Principle}
Note that the SE of MS-CD-MAMP in \eqref{Eqn:MAMPSE2} is {multi-dimensional and thus complicated}, making it difficult to {be used} directly for analyzing the achievable rate. To address this problem, we can obtain the following lemma to simplify the multi-dimensional SE analysis of MS-CD-MAMP by using the scalar SE of MS-CD-OAMP, similar to \cite[Lemma 3]{CodeMAMP2023}.

\begin{lemma}[Fixed-Point Consistency]\label{lem:fixedpoint}
Let the SE fixed point of MS-CD-MAMP in \eqref{Eqn:MAMPSE2} be $(v_{*}^{\gamma}, \tilde{v}_{*}^{\phi})$, where $v_{*}^{\gamma}=\lim\limits_{\ell \to \infty } v_{\ell,\ell}^{\gamma}$ and $v_{*}^{\phi}=\lim\limits_{\ell \to \infty } \tilde{v}_{\ell,\ell}^{\phi}$. MS-CD-MAMP and MS-CD-OAMP/VAMP can achieve the same SE fixed point $(v_{*}^{\gamma}, \tilde{v}_{*}^{\phi})$ for arbitrary fixed Lipschitz-continuous $\hat{\phi}_{\ell}(\cdot)$.
\end{lemma}
\begin{IEEEproof}
Based on \eqref{Eqn:rev_mimo}, the received signal in $\mathcal{T}$ time slots is rewritten as 
\BE\label{Eqn:allrev}
\bf{Y}=\bf{H}\bf{X}+\bf{W},
\EE
where $\bf{Y}=\text{diag}\{\bf{y}_1, ..., \bf{y}_{\mathcal{T}}\}$, $\bf{H}=\text{diag}\{\bf{H}_1, ..., \bf{H}_{\mathcal{T}}\}$,  $\bf{X}=\text{diag}\{\bf{x}_1, ..., \bf{x}_{\mathcal{T}}\}$, and $\bf{W}=\text{diag}\{\bf{w}_1, ..., \bf{w}_{\mathcal{T}}\}$. Based on \eqref{Eqn:allrev}, the fixed-point consistency of MS-CD-MAMP and MS-CD-OAMP holds for $\mathcal{T}$ time slots according to \cite[Lemma 3]{CodeMAMP2023}.

\end{IEEEproof}

\begin{figure}[t]
\centering  
\subfigure[Equivalent MS-CD-MAMP receiver: $\eta_{\ell}(\cdot)$ denotes the enhanced TDD consisting of $\{\hat{\gamma}_{t, \ell}(\cdot)\}$, damping, orthogonal operations, IUT and UT operations. $\hat{\phi}_{\ell}(\cdot)$ denotes the demodulation and APP decoder.]{
\includegraphics[width=0.8\columnwidth]{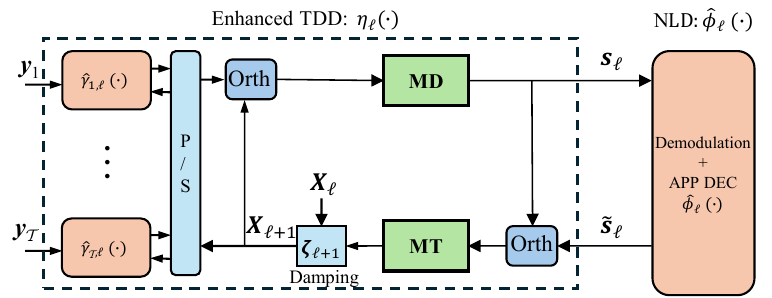}\label{Fig:systemModel2}}
\centering 
\subfigure[Variational SE: $\eta_{\text{SE}}$ and $\hat{\phi}_{\text{SE}}^{\mathcal{C}}$ are the variational signal-to-interference-plus-noise ratio (SINR) and MSE transfer functions of $\eta_{\ell}(\cdot)$ and $\hat{\phi}_{\ell}(\cdot)$, respectively.]{ 
\includegraphics[width=0.65\columnwidth]{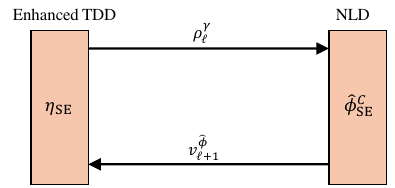}\label{Fig:SE2}
}
\caption{Equivalent MS-CD-MAMP receiver and variational state evolution.} \label{Fig:MAMPv2}
\vspace{-0.5cm}
\end{figure}

On the other hand, it is important to note that in \eqref{Eqn:nld}, NLD $\phi_{\ell}(\cdot)$ is composed of  $\hat{\phi}_{\ell}(\cdot)$ (MMSE demodulation and APP decoding) and orthogonalization, making it challenging to directly analyze the achievable rate of MS-CD-MAMP using the I-MMSE lemma in \cite{GuoTIT2005}. As a result, all orthogonalizations, IUT, and UT transformations are incorporated into the TDD $\gamma_{\ell}(\cdot)$ to constitute an enhanced TDD $\eta_{\ell}(\cdot)$. 
As shown in Fig.~\ref{Fig:systemModel2},  the equivalent MS-CD-MAMP is given by
\BE\label{Eqn:MAMPSE3}
\begin{aligned}  
\text{Enhanced TDD:}&\quad \bf{s}_{\ell}=\eta_{\ell}\left(\tilde{\bf{s}}_{\ell}\right),\\
\text{NLD:}&\quad \tilde{\bf{s}}_{{\ell}+1} =\hat{\phi}_{\ell}\left(\bf{s}_{\ell}\right).
\end{aligned}  
\EE

It is worth noting that this equivalent transformation in \eqref{Eqn:MAMPSE3} does not change the SE fixed point (i.e., converged performance) of MS-CD-MAMP. {For simplicity, the equivalent MS-CD-MAMP in Fig. \ref{Fig:systemModel2} is simply known as MS-CD-MAMP.} Based on Lemma~\ref{lem:fixedpoint} and SE of OAMP in \cite{LeiOptOAMP,YuhaoTcom2022},  as shown in Fig.~\ref{Fig:SE2}, a SISO variational SE (VSE) of MS-CD-MAMP is given in the following lemma, which  significantly simplifies the achievable rate analysis and optimal code design.


\begin{lemma}[VSE of MS-CD-MAMP]\label{lem:VSE_MAMP}
Let $\rho_{\ell}^{\gamma} = 1/ v_{\ell,\ell}^{\gamma}$ and $v^{\hat{\phi}}_{\ell}\equiv\tfrac{1}{M}\mr{E}\{\left \| \tilde{\bf{s}}_{\ell} -\bf{s} \right \|^2\}$ denote the input signal-to-interference-plus-noise ratio (SINR) and the output MSE of $\hat{\phi}_{\ell}(\cdot)$, respectively. The VSE transfer functions of MS-CD-MAMP can be represented as
\BS\label{Eqn:VSE_MAMP}
\begin{align}
    \!\!\!\!\text{TDD}: \; & \rho^{\gamma}_{\ell}= \eta_{\rm{SE}}(v^{\hat{\phi}}_{\ell}) = (v^{\hat{\phi}}_{\ell})^{-1}\!-\!\left[\frac{1}{\mathcal{T}}\sum_{t=1}^{\mathcal{T}}\hat{\eta}_{t,\mr{SE}}^{-1}(v^{\hat{\phi}}_{\ell})\right]^{-1} ,\label{Eqn:VSE-MLD}\\
     \!\!\!\!\mathrm{NLD}:\;& v^{\hat{\phi}}_{\ell+1} = \hat{\phi}^{\mathcal{C}}_{\rm{SE}}(\rho^{\gamma}_{\ell}) = \mr{mmse} \left\{\bf{s}|\sqrt{\rho^{\gamma}_{\ell}}\bf{s}+\bf{z}, \Phi_{\mathcal{C}}\right\},\label{Eqn:VSE-NLD}
\end{align}
\ES
where $\hat{\eta}_{t,\mr{SE}}(v)= \tfrac{1}{NJ}{\rm{tr}}\{[snr\bf{H}_t^{\rm{H}}\bf{H}_t+v^{-1}\bf{I}]^{-1}\}$ denotes the MSE function of LMMSE detector in OAMP\cite{LeiOptOAMP,YuhaoTcom2022}, $\hat{\eta}_{t,\mr{SE}}^{-1}(\cdot)$ is the inverse of $\hat{\eta}_{t,\mr{SE}}(\cdot)$, $\bf{z}\sim \mathcal{CN}(\bf{0}, \bf{I})$ is an AWGN vector independent of $\bf{s}$, and $\Phi_{\mathcal{C}}$ denotes the coding constraint, i.e., $\bf{s} \in {\mathcal{C}}$ ($\mathcal{C}$ is the set of FEC codewords).
{
\begin{IEEEproof}
    See Appendix.
\end{IEEEproof}
}
\end{lemma}

Although VSE in \eqref{Eqn:VSE_MAMP} cannot accurately predict the MSE performance of MS-CD-MAMP at each iteration, it can converge to the same fixed point as the original SE in \eqref{Eqn:MAMPSE2}, thereby enabling the achievable rate analysis and optimal coding principle.

The following two lemmas present two tight upper bounds on the MMSE decoding function.
\begin{lemma}[Decoding Gain]\label{lem:codegain}
The demodulation transfer function $\hat{\phi}_{\mr{SE}}^{\mathcal{S}}(\cdot)$ is the upper bound of the decoding transfer function $\hat{\phi}_{\mr{SE}}^{\mathcal{C}}(\cdot)$ due to the coding gain, i.e.,
\BE\label{Eqn:CodeMAMP1}
\hat{\phi}^{\mathcal{C}}_{\rm{SE}}(\rho^{\gamma}_{\ell}) < \hat{\phi}^{\mathcal{S}}_{\rm{SE}}(\rho^{\gamma}_{\ell}),  \quad \mr{for}\;\; 0\leq \rho^{\gamma}_{\ell} \leq snr,	
\EE
\end{lemma}
where $\hat{\phi}^{\mathcal{S}}_{\rm{SE}}(\rho^{\gamma}_{\ell}) = \mr{mmse} \{\bf{s}|\sqrt{\rho^{\gamma}_{\ell}}\bf{s}+\bf{z}, \Phi_{\mathcal{S}}\}$ with modulate constraint $\Phi_{\mathcal{S}}$, i.e., $s_i \sim P_S(s_i)$.

\begin{figure}[!tbp]
	\centering
	\includegraphics[width=0.75\columnwidth]{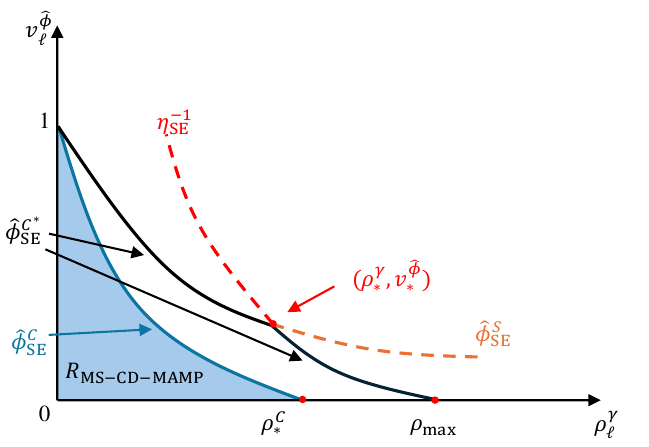}\vspace{-0.3cm}
	\caption{The VSE transfer functions of MS-CD-MAMP, where $\hat{\phi}^{\mathcal{C}^*}_{\rm{SE}}$ is the optimal MMSE function of decoder,
 $\eta^{-1}_{\rm{SE}}$ is the inverse function of $\eta_{\rm{SE}}$, and the MMSE functions of constellation and code constraint in NLD are indicated individually by $\hat{\phi}^{\mathcal{S}}_{\rm{SE}}$ and $\hat{\phi}^{\mathcal{C}}_{\rm{SE}}$ with $\hat{\phi}_{\mr{SE}}^{\mathcal{C}}(\rho_{*}^{\mathcal{C}})=0$. $(\rho_*^{\gamma}, v_*^{\hat{\phi}})$ is the VSE fixed point between $\eta^{-1}_{\rm{SE}}$ and $\hat{\phi}^{\mathcal{S}}_{\rm{SE}}$. }\label{Fig:SEcurve}
\end{figure}

As shown in  Fig.~\ref{Fig:SEcurve}, assuming that there exists a single fixed point $(\rho_*^{\gamma}, v_*^{\hat{\phi}})$ between $\eta_{\rm{SE}}^{-1}(\cdot)$ and $\hat{\phi}^{\mathcal{S}}_{\rm{SE}}(\cdot)$, the converged performance of MS-CD-MAMP is not error-free due to $v_*^{\hat{\phi}}>0$. Therefore, to successfully recover the message at the receiver, it is necessary to guarantee an available decoding tunnel between the decoder and the detector, according to the iterative detection principle. Consequently, the following lemma presents the error-free decoding condition.




\begin{lemma}[Error-Free Decoding]\label{lem:err_free}
For given  $\hat{\phi}_{\mr{SE}}^{\mathcal{C}}(\cdot)$, the MS-CD-MAMP receiver can achieve error-free decoding if and only if 
\BE\label{Eqn:CodeMAMP2}
    \hat{\phi}^{\mathcal{C}}_{\rm{SE}}(\rho^{\gamma}_{\ell}) < \eta_{\rm{SE}}^{-1}(\rho^{\gamma}_{\ell}), \quad \mr{for}\;\; 0 \leq \rho^{\gamma}_{\ell} \leq snr.
\EE
\end{lemma}


Therefore, we derive the upper bounds of $\hat{\phi}_{\mr{SE}}^{\mathcal{C}}(\cdot)$ for decoder based on Lemma~\ref{lem:codegain} and Lemma~\ref{lem:err_free} as following,
\BE\label{Eqn:CodeMAMP3}
\!\!\!\hat{\phi}^{\mathcal{C}}_{\rm{SE}}(\rho^{\gamma}_{\ell}) <  \min\{\hat{\phi}^{\mathcal{S}}_{\rm{SE}}(\rho^{\gamma}_{\ell}), \eta_{\rm{SE}}^{-1}(\rho^{\gamma}_{\ell})\},  \; \mr{for}\; 0\leq \rho^{\gamma}_{\ell} \leq snr.
\EE


\LL{Then, based on \eqref{Eqn:VSE_MAMP}, \eqref{Eqn:CodeMAMP3}, and I-MMSE lemma\cite{GuoTIT2005}, the achievable rate of MS-CD-MAMP is calculated according to the following lemma, assuming an infinite iterations.
\begin{lemma}[Achievable Rate of MS-CD-MAMP]\label{lem:Rate}
The average achievable rate of MS-CD-MAMP per transmit antenna per transmitted symbol with fixed $\hat{\phi}_{\mr{SE}}^{\mathcal{C}}(\cdot)$ is
\BE \label{Eqn:rate_MAMP}
\begin{aligned}
 &R_{\text{MS-CD-MAMP}} = \int_{0}^{\rho_{*}^{\mathcal{C}}}
 \hat{\phi}_{\mr{SE}}^{\mathcal{C}}(\rho_{\ell}^{\gamma}) d \rho_{\ell}^{\gamma},   \\
 &\begin{array}{l@{\quad}l}
 {\rm s.t.} &  \hat{\phi}_{\mr{SE}}^{\mathcal{C}}(\rho_{\ell}^{\gamma})< \min\{\hat{\phi}^{\mathcal{S}}_{\rm{SE}}(\rho^{\gamma}_{\ell}), \eta_{\rm{SE}}^{-1}(\rho^{\gamma}_{\ell})\}, \\
 &0\le \rho_{*}^{\mathcal{C}} \le \rho_{\mr{max}},
 \end{array}
\end{aligned}
\EE    
where $\rho_{*}^{\mathcal{C}}=\hat{\phi}_{\mr{SE}}^{\mathcal{C}^{-1}}(0)$ and $\rho_{\mr{max}}=\eta^{-1}_{\mr{SE}}(0)$.
\end{lemma}
}

To maximize the achievable rate of MS-CD-MAMP receiver while ensuring error-free decoding, the MMSE function $\hat{\phi}_{\mr{SE}}^{\mathcal{C}^*}(\cdot)$ of the optimal decoder is given in the following lemma, according to \eqref{Eqn:rate_MAMP}.
\begin{lemma}[Optimal Code Design]\label{lem:optimal code design}
    The optimal coding principle of MS-CD-MAMP is
    \BE\label{Eqn: Optimal coded desdign}
    \begin{aligned}
    &\hat{\phi}_{\mr{SE}}^{\mathcal{C}}(\rho_{\ell}^{\gamma}) \to \hat{\phi}_{\mr{SE}}^{\mathcal{C}^*}(\rho_{\ell}^{\gamma}), \\
     {\rm s.t.}\;&\hat{\phi}_{\mr{SE}}^{\mathcal{C}^*}(\rho_{\ell}^{\gamma})=\mr{min}\{\hat{\phi}^{\mathcal{S}}_{\rm{SE}}(\rho_{\ell}^{\gamma}), \eta_{\rm{SE}}^{-1}(\rho_{\ell}^{\gamma})\},\\
    &0\le \rho_{\ell}^{\gamma} \le \rho_{\mr{max}}.
    \end{aligned}
    \EE    
\end{lemma}
\LL{Based on Lemma \ref{lem:optimal code design}, the time-domain channel parameters affect $\eta_{\rm{SE}}^{-1}(\cdot)$, while the coding parameters and constellation mapping affect 
$\hat{\phi}^{\mathcal{S}}_{\rm{SE}}(\cdot)$, which jointly determines the optimal decoding curve and thereby influences code design.
}

The following theorem directly derives the maximum achievable rate of MS-CD-MAMP based on Lemma~\ref{lem:Rate} and Lemma~\ref{lem:optimal code design}.

\begin{theorem}[Maximum Achievable Rate of MS-CD-MAMP]\label{The:maxrate}
The average maximum achievable rate of MS-CD-MAMP per transmit antenna per transmitted symbol is 
\BE\label{Eqn:rateMAMP}
  	R_{\rm{MS-CD-MAMP}}^{\rm{max}} \to  \int_{0}^{\rho_{\rm{max}}}  \hat{\phi}^{\mathcal{C}^*}_{\rm{SE}}(\rho_{\ell}^{\gamma}) d\rho_{\ell}^{\gamma},
\EE
where $\hat{\phi}^{\mathcal{C}^*}_{\rm{SE}}(\rho_{\ell}^{\gamma})={\rm{min}}\{\hat{\phi}^{\mathcal{S}}_{\rm{SE}}(\rho_{\ell}^{\gamma}), \eta_{\rm{SE}}^{-1}(\rho_{\ell}^{\gamma})\}$.
\end{theorem}

\LL{
Based on Theorem~\ref{The:maxrate}, the total transmission rate of all antennas is $R_{\rm{MS-CD-MAMP}}^{\rm{total}}=JR_{\rm{MS-CD-MAMP}}^{\rm{max}}$.
}


In addition, based on Theorem~\ref{The:MSE_eq} and Lemma~\ref{lem:fixedpoint}, the VSE of MS-CD-MAMP is still held for symbol domain MAMP in \cite{MAMPOTFSconf} for unitary matrix transform-based multicarrier modulations; thus, the following corollary can be deduced straightly.

\begin{corollary}[Achievable Rate Equivalency of MS-CD-MAMP and Symbol-Domain MAMP in \cite{MAMPOTFSconf}]\label{cor:rateMAMP}
For unitary matrix transform-based multicarrier modulations, the VSEs of the MS-CD-MAMP and symbol-domain MAMP are the same, thereby achieving the same achievable rates with the same optimal code.
\end{corollary}

\begin{corollary}[Achievable Rate Equivalency of Unitary Matrix Transform-Based Multicarrier Modulation with MS-CD-MAMP]\label{cor:rateMAMP2} For multicarrier modulations based on the unitary transform matrix, such as OFDM, OTFS, and AFDM, their VSEs of MS-CD-MAMP are the same, as are their achievable rates, according to Theorem~\ref{The:maxrate}.
\end{corollary}

\LL{It should be noted that under multi-dimensional constraints, such as channel matrix distributions and arbitrary discrete inputs, directly calculating the channel mutual information of MIMO multicarrier systems is NP-hard, which could potentially be addressed using the replica method. In contrast, the multi-slot SE can accurately predict the asymptotic performance of the MS-CD-MAMP, and codes optimized based on the SE achieve performance close to the theoretical limits, indicating that MS-CD-MAMP achieves MMSE optimality. Therefore, based on the I-MMSE lemma \cite{GuoTIT2005}, we conjecture that the achievable rate of MS-CD-MAMP equals the channel mutual information, i.e., MS-CD-MAMP is constrained-capacity optimal.
\begin{conjecture}[Constrained Capacity Optimality of MS-CD-MAMP]
Assume there is a unique SE fixed point in \eqref{Eqn:MAMPSE}, the MS-CD-MAMP is constrained-capacity optimal in MIMO multicarrier systems, i.e., the total transmission rate $R_{\rm{MS-CD-MAMP}}^{\rm{total}}$ is equal to the constrained capacity of MIMO multicarrier systems.
\end{conjecture}
}

\begin{figure*}[t!]
	\centering %
	\includegraphics[width=0.85\textwidth]{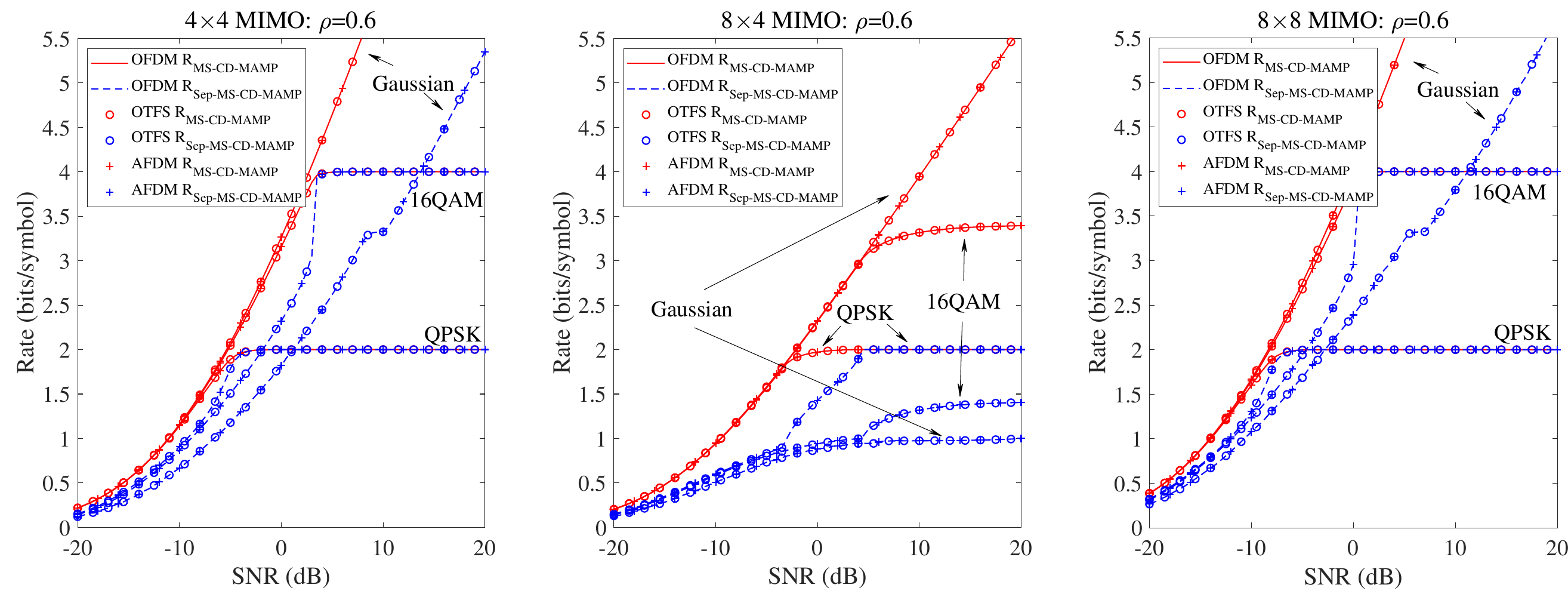}\vspace{-0.1cm}
	\caption{Maximum achievable rates  of OFDM/OTFS/AFDM in {$4\times4$, $8\times4$, and $8\times8$ MIMO} with {$\rho_{tx}=\rho_{rx}=\rho=0.6$}, where $P_{u,j}=5$, $v=500$km/h and \{Gaussian, 16QAM, QPSK\} signaling,  $\mathcal{T}=5000, N=256$ for AFDM and OFDM, and $K=8, L=32$ for OTFS.}\label{Fig:rateModulation}  
\end{figure*}
\begin{figure}[t] 
\centering  
\subfigure[Comparison for different {velocities} with $P_{u,j}=5$.]{
\includegraphics[width=1\columnwidth]{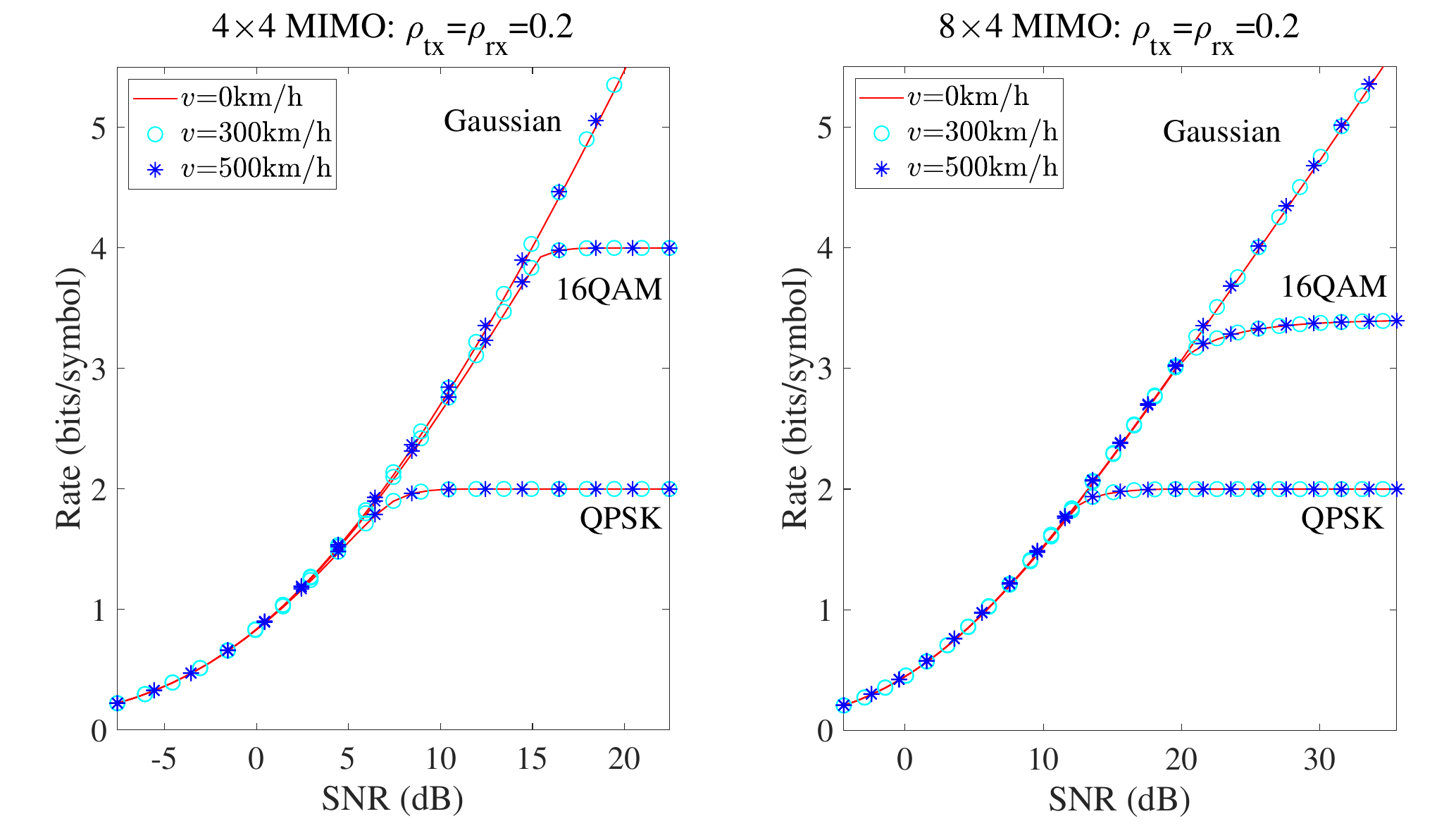}\label{Fig:MIMOspeed}}
\centering 
\subfigure[Comparison for different {number of multipaths} with $v=500$km/h.]{ 
\includegraphics[width=1\columnwidth]{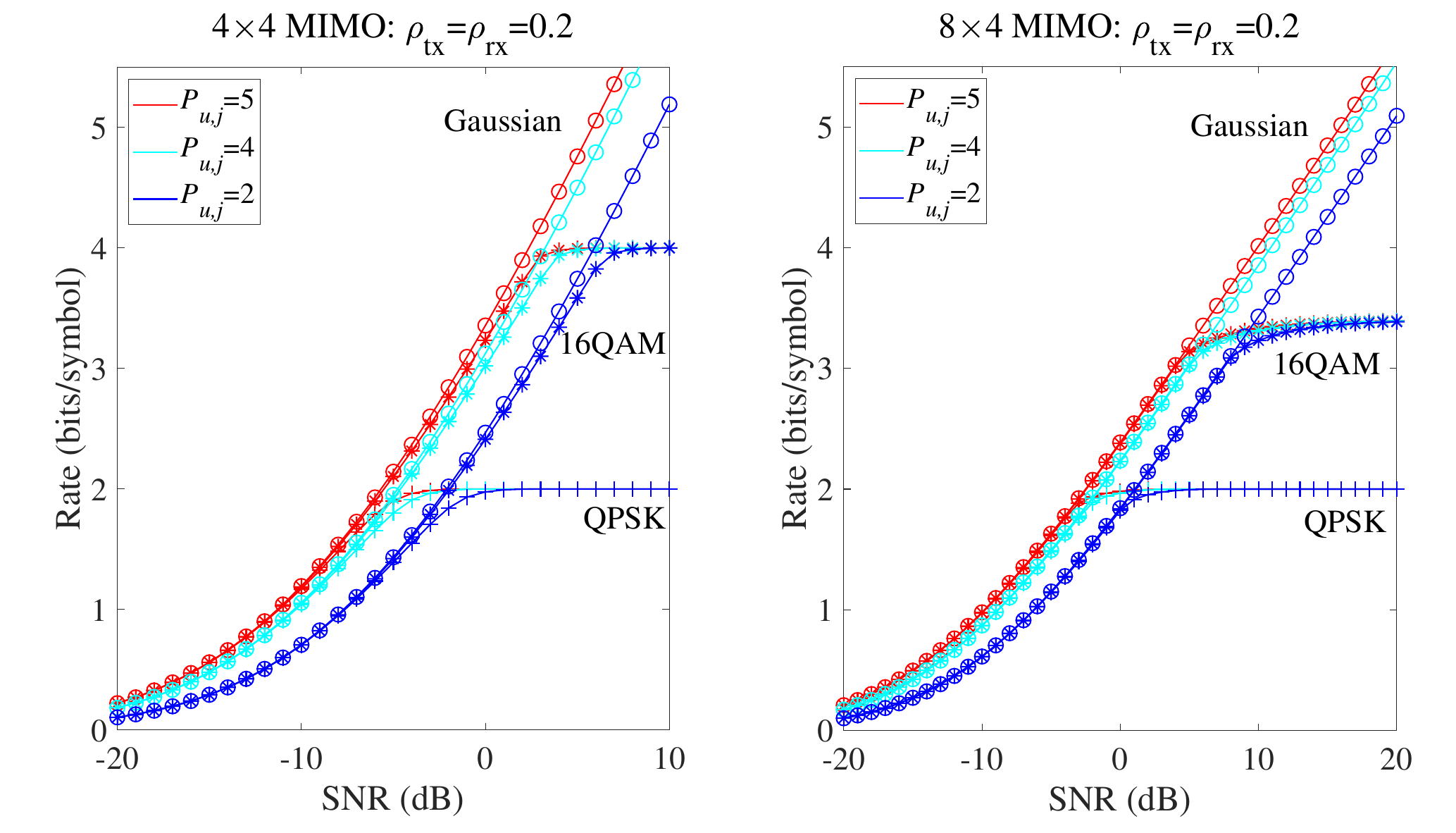}\label{Fig:multipath}
}
\caption{Maximum achievable rate comparison for {different} number of multipaths and {user velocities} at $4\times4$ and $8\times4$ MIMO systems with {$\rho=0.2$}, where $P_{u,j}\in\{2,4,5\}$, $v \in \{0,300,500\}$km/h, $\mathcal{T}=5000, N=256$, \{Gaussian, 16QAM, QPSK\} signaling.} \label{Fig:rateSpeedPath}
\end{figure}

\subsection{Maximum Achievable Rates Comparisons of MS-CD-MAMP in MIMO-OFDM/OTFS/AFDM}\label{sec:correlatedMIMO}
Based on Theorem \ref{The:maxrate}, we evaluate the maximum average achievable rates of the MS-CD-MAMP receiver in MIMO-OFDM/OTFS/AFDM with different input signaling and parameter configurations as follows:
    
\emph{1}) \emph{Maximum achievable rate comparison for different modulations:} Fig.~\ref{Fig:rateModulation} illustrates the maximum achievable rates of MS-CD-MAMP receiver in different MIMO-multicarrier systems (i.e., OFDM, OTFS, and AFDM) with $ N=256$, $\mathcal{T}=5000$\footnote{\LL{Based on \eqref{Eqn:VSE_MAMP},  we set  $\mathcal{T}=5000$  and average the achievable rates over all generated channel realizations to accurately capture its statistical distribution.}},$v=500$km/h, and different signaling (i.e., QPSK, 16QAM and Gaussian signaling), where antenna configuration
    $(J, U)=(4, 4), (8, 4)$, and $(8, 8)$ with $\rho_{\rm{tx}}=\rho_{\rm{rx}}=\rho=0.6$, respectively. It is shown that the achievable rates of MS-CD-MAMP receiver with different multicarrier modulations completely overlap in the available SNR region. The reason is that the VSE of MS-CD-MAMP is the same for different multicarrier modulations with a unitary transform matrix, as stated in Lemma \ref{lem:VSE_MAMP}, ensuring that the maximum achievable rate is also the same based on Theorem \ref{The:maxrate}. Therefore, in the subsequent achievable rate analysis, we do not distinguish between specific types of multicarrier modulations.
    
\emph{2}) \emph{Maximum achievable rate comparison with existing methods:} A separable MS-CD-MAMP receiver, denoted as Sep-MS-CD-MAMP, is {considered similar to} \cite{OTFS-OAMP,DD-OAMP,MAMPOTFSconf}, in which the full iteration {is applied} between the detection and nonlinear demodulation with separate APP decoding. Fig.~\ref{Fig:SEcurve} shows that the maximum achievable rate of Sep-MS-CD-MAMP is $\int_{0}^{\rho_{*}^{\gamma}}  \hat{\phi}^{\mathcal{S}}_{\rm{SE}}(\rho_{\ell}^{\gamma}) d\rho_{\ell}^{\gamma}$. As shown in Fig.~\ref{Fig:rateModulation},  the achievable rates of MS-CD-MAMP are higher than those of Sep-MS-CD-MAMP for arbitrary input signaling, especially at low to medium SNR levels and with Gaussian signaling. With QPSK signaling at rate $1$, MS-CD-MAMP achieves approximately a $2$ dB gain over Sep-MS-CD-MAMP in $4\times 4$ and $8\times8$ MIMO and a $7$~dB gain in $8\times4$ MIMO.
    This is because of the inherently poor demodulation performance of higher-order signaling and the losses of the joint iterative gain with the APP decoder. As a result, the rate loss of Bayes-optimal MS-CD-MAMP with ideal point-to-point (P2P) capacity-achieving codes is $R_{\rm{loss}}=\int_{\rho_{*}^{\gamma}}^{\rho_{\rm{max}}}  \hat{\phi}^{\mathcal{C}^*}_{\rm{SE}}(\rho_{\ell}^{\gamma}) d\rho_{\ell}^{\gamma}$. 

\begin{figure*}[t!]
	\centering 
	\includegraphics[width=1\textwidth]{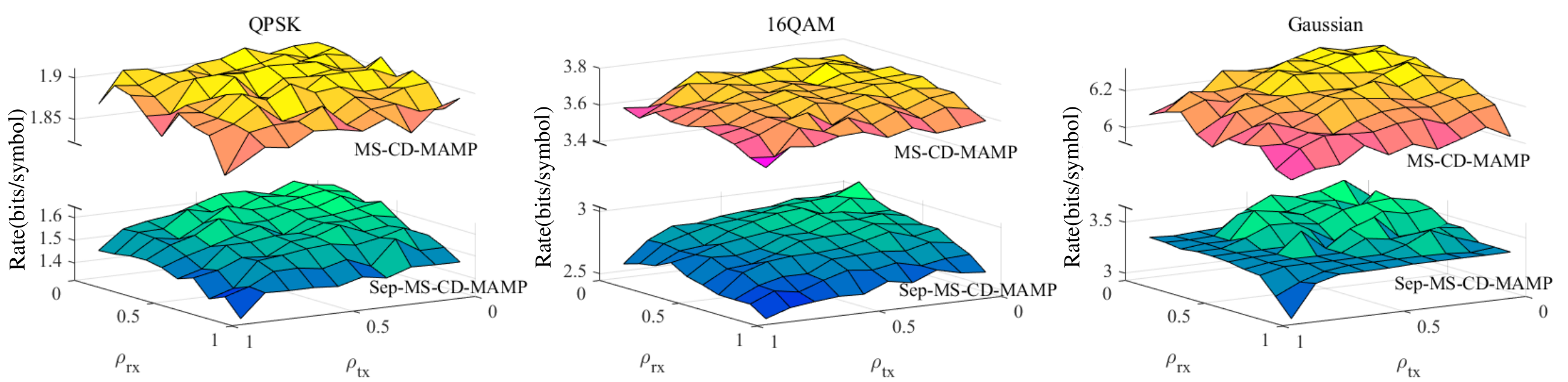}
	\caption{Maximum achievable rates of MS-CD-MAMP with different antenna correlations in $4\times 4$ MIMO, where $(\rho_{\text{tx}},\rho_{\text{rx}})\in\{0,0.1,\cdots,0.9\}$,  $P_{u,j}=5$, $v=500$km/h, $\mathcal{T}=5000, N=256$, QPSK signaling at {SNR = $6.4$ dB}, 16QAM signaling at SNR = $14.4$ dB, and Gaussian signaling at SNR = $22.4$ dB.}\label{Fig:rateRho}  
\end{figure*}

\emph{3})  \emph{Maximum achievable rate comparison for different device velocities:} It is well known that different velocities cause different Doppler effects on the system. Therefore, Fig.~\ref{Fig:MIMOspeed} presents the achievable rate comparison of MS-CD-MAMP at velocity $v=\{0,300,500\}$~km/h in MIMO with QPSK, 16QAM, and Gaussian signaling, where $\mathcal{T}=5000, N=256$, 
    $(J,U)=(4,4)$ and $(8,4)$, {$\rho=0.2$}. Surprisingly, the maximum achievable rate of MS-CD-MAMP is almost the same for different velocities. This indicates that the {proposed} optimal coding scheme is robust to different mobile velocity scenarios without the need for re-optimization.
    
\emph{4}) \emph{Maximum achievable rate comparison for different  number of multipaths:} Since different number of multipaths directly affect the {degrees of freedom in the channel}, Fig.~\ref{Fig:multipath} presents the achievable rate comparisons of MS-CD-MAMP in MIMO with different number of multipaths, where input signaling is QPSK, 16QAM, Gaussian, 
    $(J,U)=(4,4)$ and $(8,4)$, {$\rho=0.2$}, $v=500$km/h. It is clearly seen that the achievable rates of MS-CD-MAMP increase as the number of multipaths increases.

\emph{5})  \emph{Maximum achievable rate comparison for different antenna correlations:} {In practical MIMO scenarios, different correlations exist between transmit and receive antennas due to the varying types and locations of transmit devices and base stations. Fig. \ref{Fig:rateRho} shows the maximum achievable rates of MS-CD-MAMP for different input signaling (i.e., QPSK, 16QAM, and Gaussian) in MIMO systems with $(J, U)=(4,4)$ and the correlation coefficients $(\rho_{\rm{tx}},\rho_{\rm{rx}}) \in \{0, 0.1,\cdots,0.9\}$.  It can be observed that the achievable rate of MS-CD-MAMP decreases slightly and monotonically as the correlation parameter increases, whereas the achievable rate of Sep-MS-CD-MAMP degrades more significantly. Moreover, across all antenna correlation levels, the achievable rate gains of MS-CD-MAMP over Sep-MS-CD-MAMP progressively increase with higher modulation orders. These results highlight the crucial role of optimal coding in the iterative detection process of the MS-CD-MAMP receiver. } 
\PZY{
}

\vspace{-0.3cm}
\section{Numerical Results}\label{sec:NR}
In this section, we provide the BER performance of optimized finite-length LDPC codes for MS-CD-MAMP receiver in MIMO-multicarrier modulation systems. Meanwhile, BER comparisons with existing benchmark schemes are presented.

\subsection{Simulation Configuration}
We consider the design of LDPC code  in MIMO-multicarrier modulation systems with QPSK signaling, where the antenna configuration $(J, U) \in \{(4\times 4), (8\times 4), (8 \times 8)\}$. For simplicity,  the correlated parameters of MIMO channel are set as $\rho=\rho_{\rm{tx}}=\rho_{\rm{rx}}=\{0.2,0.6\}$. The sizes of modulation matrix are $N=256$ for OFDM and AFDM, and $K=8$, $L=32$ for OTFS, respectively. $\mathcal{T}\in\{1, 5, 25\}$, the carrier frequency is  $4$ GHz, and subcarrier interval $\Delta f =15\rm{kHz}$, where $P_{u,j} \in \{2, 5\}$, the velocity of the device $v \in \{0, 300, 500\}$~km/h, and the RRC rolloff factor is set to $0.4$ for the transceiver.




 
\begin{table}[b]\scriptsize
\caption{Optimized LDPC codes for MS-CD-MAMP in MIMO multicarrier modulation systems.}\label{Tab:LDPCparam}
\centering
\begin{adjustbox}{width=\linewidth}
\begin{tabular}{|c|cccccc|}
\hline
Channel Types                       & \multicolumn{6}{c|}{Correlated MIMO channel}                                                                                                                                                                                                                                                                                                                                                           \\ \hline
$N$                                 & \multicolumn{6}{c|}{256}                                                                                                                                                                                                                                                                                                                                                                               \\ \hline
Antennas Configuration              & \multicolumn{1}{c|}{$4\times4$}                                    & \multicolumn{3}{c|}{$8\times4$}                                                                                                                                                                              & \multicolumn{2}{c|}{$8\times8$}                                                                                    \\ \hline
$\rho_{\text{tx}}=\rho_{\text{rx}}$ & \multicolumn{3}{c|}{0.6}                                                                                                                                                                                     & \multicolumn{1}{c|}{0.2}                                           & \multicolumn{1}{c|}{0.6}                                           & 0.2                                           \\ \hline
$P_{u,j}$                           & \multicolumn{1}{c|}{5}                                             & \multicolumn{1}{c|}{2}                                             & \multicolumn{1}{c|}{5}                                             & \multicolumn{1}{c|}{5}                                             & \multicolumn{1}{c|}{5}                                             & 5                                             \\ \hline
$v$                                 & \multicolumn{6}{c|}{$\{0, 300, 500\}$ km/h}                                                                                                                                                                                                                                                                                                                                                            \\ \hline
Codeword length                     & \multicolumn{6}{c|}{102400}                                                                                                                                                                                                                                                                                                                                                                            \\ \hline
$R_{\rm{LDPC}}$                     & \multicolumn{1}{c|}{0.5071}                                        & \multicolumn{1}{c|}{0.4942}                                        & \multicolumn{1}{c|}{0.4991}                                        & \multicolumn{1}{c|}{0.4960}                                              & \multicolumn{1}{c|}{0.4943}                                        &          \multicolumn{1}{c|}{0.5058}                                       \\ \hline
$\mu(X)$                            & \multicolumn{1}{c|}{\begin{tabular}[c]{@{}c@{}}$\mu_8=1$\end{tabular}} & \multicolumn{1}{c|}{\begin{tabular}[c]{@{}c@{}}$\mu_8=0.8$\\ $\mu_{30}=0.2$\end{tabular}} & \multicolumn{1}{c|}{\begin{tabular}[c]{@{}c@{}}$\mu_6=1$\end{tabular}} & \multicolumn{1}{c|}{\begin{tabular}[c]{@{}c@{}}$\mu_8=1$\end{tabular}} & \multicolumn{1}{c|}{\begin{tabular}[c]{@{}c@{}}$\mu_6=1$\end{tabular}} & \begin{tabular}[c]{@{}c@{}}$\mu_8=1$\end{tabular} \\ \hline
$\lambda(X)$                        & \multicolumn{1}{c|}{\begin{tabular}[c]{@{}c@{}}$\lambda_2=0.3059$\\ $\lambda_3=0.2179$\\$\lambda_{9}=0.1462$\\$\lambda_{10}=0.0503$\\$\lambda_{30}=0.0429$\\$\lambda_{35}=0.1510$\\$\lambda_{80}=0.0516$\\$\lambda_{90}=0.0342$\end{tabular}} & \multicolumn{1}{c|}{\begin{tabular}[c]{@{}c@{}}$\lambda_2=0.2556$\\ $\lambda_3=0.1946$\\$\lambda_{13}=0.0036$\\$\lambda_{14}=0.1942$\\$\lambda_{40}=0.0931$\\ $\lambda_{90}=0.0510$\\$\lambda_{100}=0.1087$\\$\lambda_{800}=0.0992$\end{tabular}} & \multicolumn{1}{c|}{\begin{tabular}[c]{@{}c@{}}$\lambda_2=0.6185$\\ $\lambda_3=0.0202$\\$\lambda_{19}=0.2104$\\$\lambda_{20}=0.1031$\\$\lambda_{90}=0.0497$\end{tabular}} & \multicolumn{1}{c|}{\begin{tabular}[c]{@{}c@{}}$\lambda_2=0.3978$\\ $\lambda_3=0.1023$\\$\lambda_{19}=0.0012$\\$\lambda_{20}=0.2653$\\$\lambda_{80}=0.0973$\\$\lambda_{300}=0.1361$\end{tabular}} & \multicolumn{1}{c|}{\begin{tabular}[c]{@{}c@{}}$\lambda_2=0.1796$\\ $\lambda_3=0.4299$\\$\lambda_{12}=0.1756$\\$\lambda_{13}=0.2799$\end{tabular}} & \begin{tabular}[c]{@{}c@{}}$\lambda_2=0.2991$\\ $\lambda_3=0.2170$\\$\lambda_{8}=0.1815$\\$\lambda_{27}=0.0086$\\$\lambda_{28}=0.1821$\\$\lambda_{70}=0.1022$\\$\lambda_{80}=0.0096$\end{tabular} \\ \hline
$\text{(SNR)}^{*}_{\rm{dB}}$        & \multicolumn{1}{c|}{1.68}                                           & \multicolumn{1}{c|}{5.8}                                         & \multicolumn{1}{c|}{2.3}                                         & \multicolumn{1}{c|}{2}                                              & \multicolumn{1}{c|}{1.61}                                        &                 \multicolumn{1}{c|}{1.4}                               \\ \hline
$\text{(Limit)}_{\rm{dB}}$          & \multicolumn{1}{c|}{1.48}                                         & \multicolumn{1}{c|}{5.75}                                            & \multicolumn{1}{c|}{2.25}                                          & \multicolumn{1}{c|}{1.95}                                              & \multicolumn{1}{c|}{1.56}                                           &       \multicolumn{1}{c|}{1.38}                                         \\ \hline
\end{tabular}
\end{adjustbox}
\end{table}

\subsection{Optimization of Irregular LDPC Codes for MS-CD-MAMP}
The optimization of LDPC codes is similar to \cite{YuhaoTcom2022,LeiTIT2021}, i.e., the degree distributions of variable and check nodes (${\lambda}(X)=\sum^{d_{v,\rm{max}}}_{i=2}\lambda_iX^{i-1}, {\mu}(X)=\sum^{d_{c,\rm{max}}}_{i=2}\mu_iX^{i-1}$) of LDPC codes is derived by solving a linear programming problem constrained by the successful iterative decoding with the aim of maximizing the achievable rate of MS-CD-MAMP receiver, where $d_{v,\rm{max}}$ 
 and $d_{v,\rm{max}}$ are the corresponding maximum degrees of variable and check nodes, respectively. The parameters of optimized LDPC codes are given in Table \ref{Tab:LDPCparam} with the target $R_{\rm{LDPC}}=0.5$ and codeword length $M=102400$, {where the theoretical limit is denotes the SNR corresponding to the maximum achievable rate of $1$.
 }

 \begin{figure*}[t]
\centering
\subfigure[Different modulations: OTFS/AFDM/OFDM.]{
\begin{minipage}[t]{0.33\textwidth}
\centering
\includegraphics[width=0.7\textwidth]{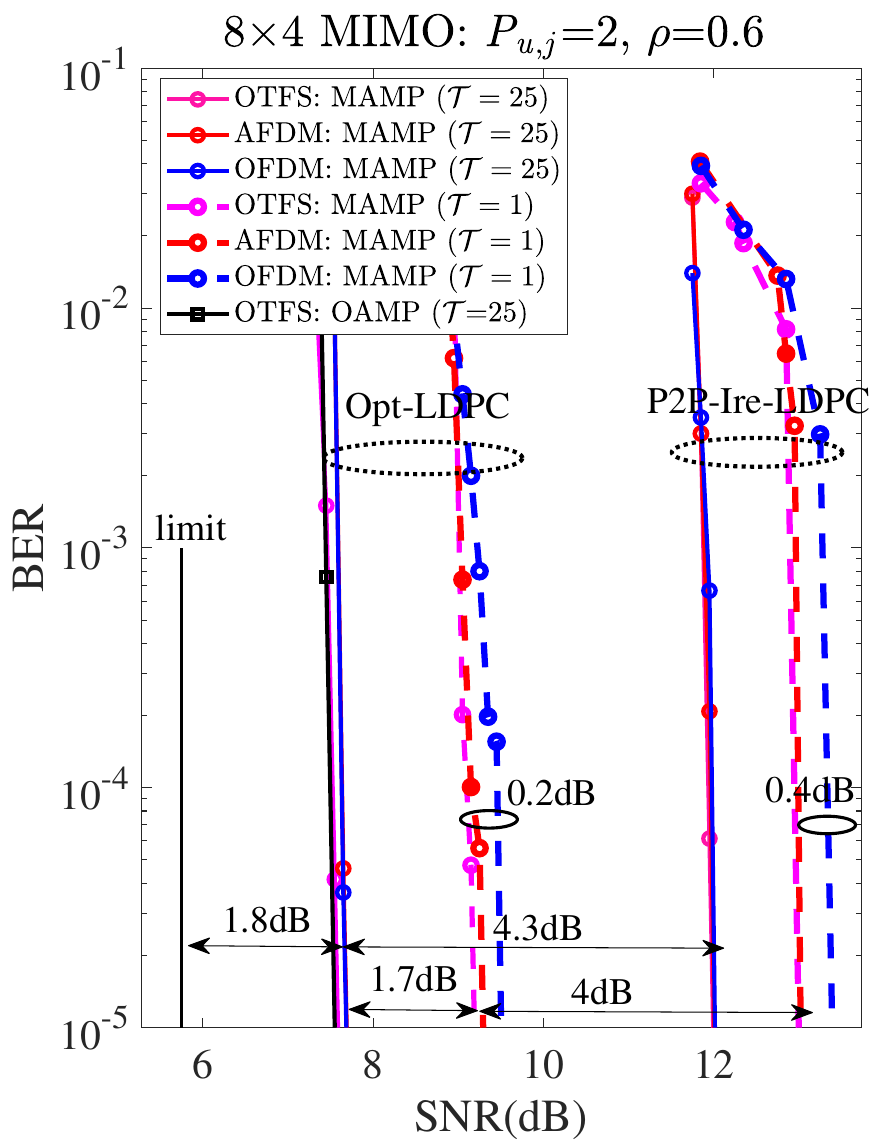}\label{fig:ber1a}
\end{minipage}%
}%
\hspace{-0.5cm}
\subfigure[Antenna correlation: $\rho \in \{0.2, 0.6\}$.]{
\begin{minipage}[t]{0.33\textwidth}
\centering
\includegraphics[width=0.7\textwidth]{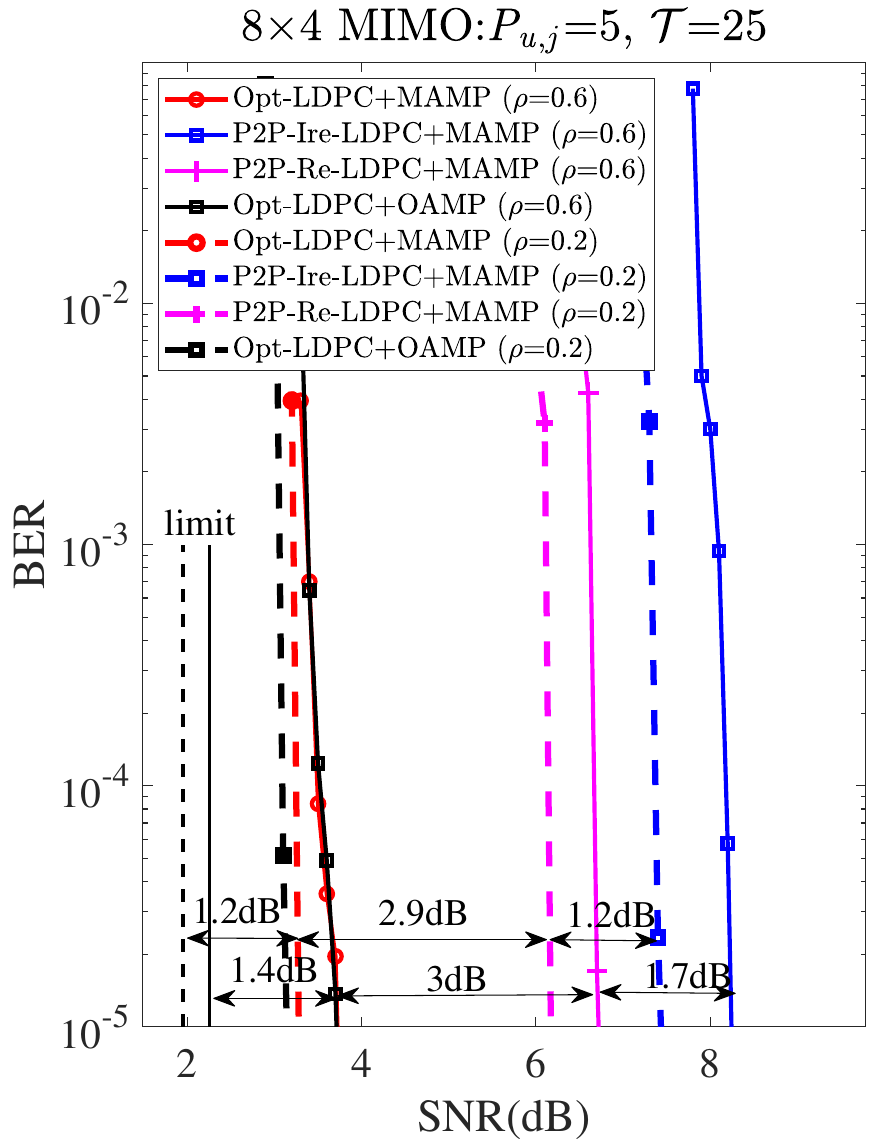}\label{fig:ber1b}
\end{minipage}%
}%
\hspace{-0.5cm}
\subfigure[$4\times4$ MIMO: $\rho=0.6$ + different $\mathcal{T}$.]{
\begin{minipage}[t]{0.33\textwidth}
\centering
\includegraphics[width=0.7\textwidth]{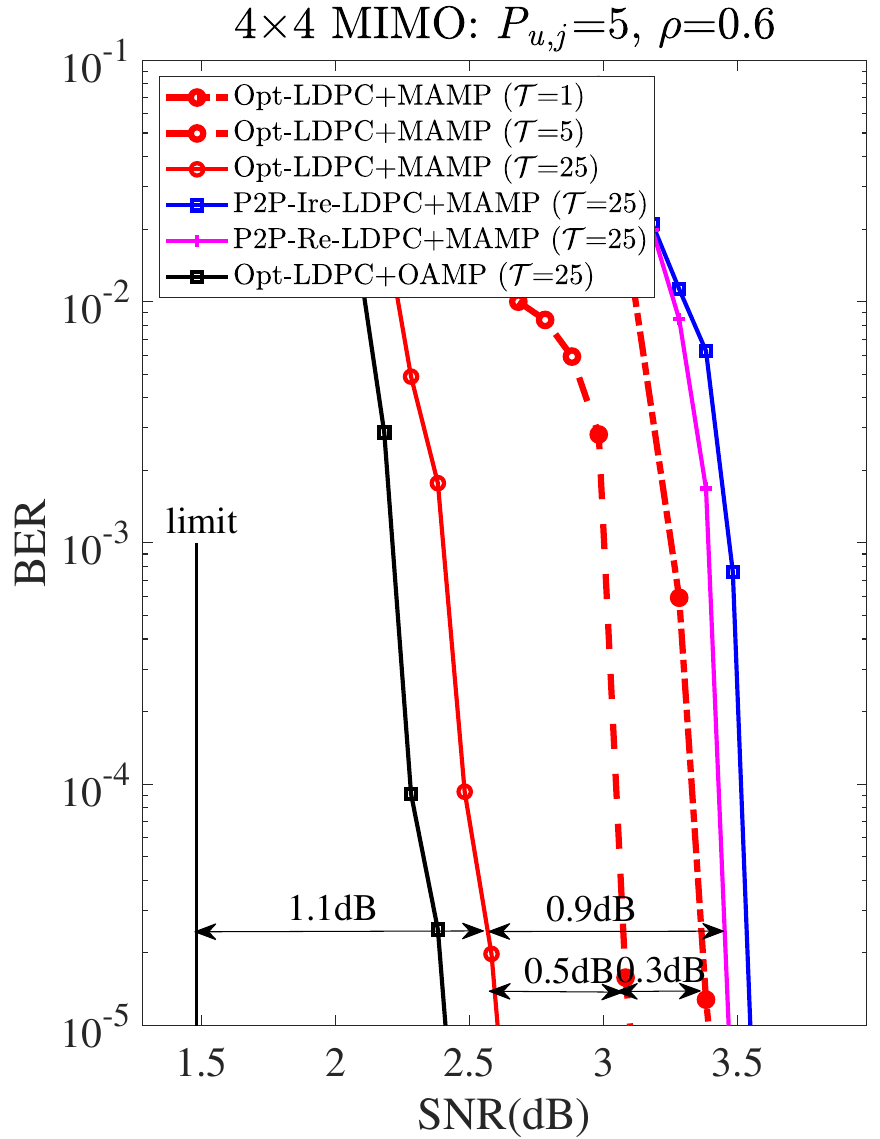}\label{fig:ber1c}
\end{minipage}
}%
\vspace{-0.3cm}
\subfigure[$8\times8$ MIMO: $\rho=0.6$ + different $\mathcal{T}$.]{
\begin{minipage}[t]{0.33\textwidth}
\centering
\includegraphics[width=0.7\textwidth]{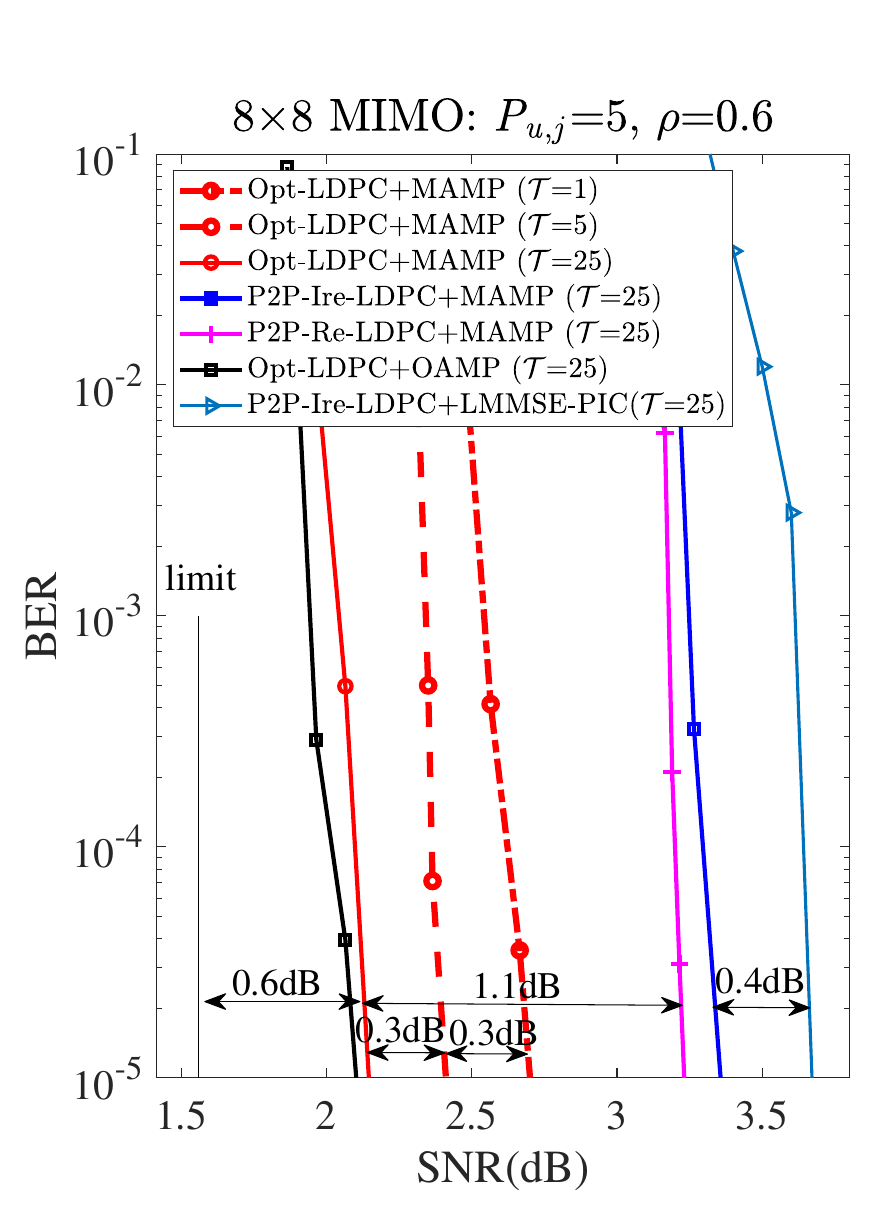}\label{fig:ber2a}
\end{minipage}
}
\hspace{-0.6cm}
\subfigure[$8\times8$ MIMO: $\rho=0.2$ + different $\mathcal{T}$.]{
\begin{minipage}[t]{0.33\textwidth}
\centering
\includegraphics[width=0.7\textwidth]{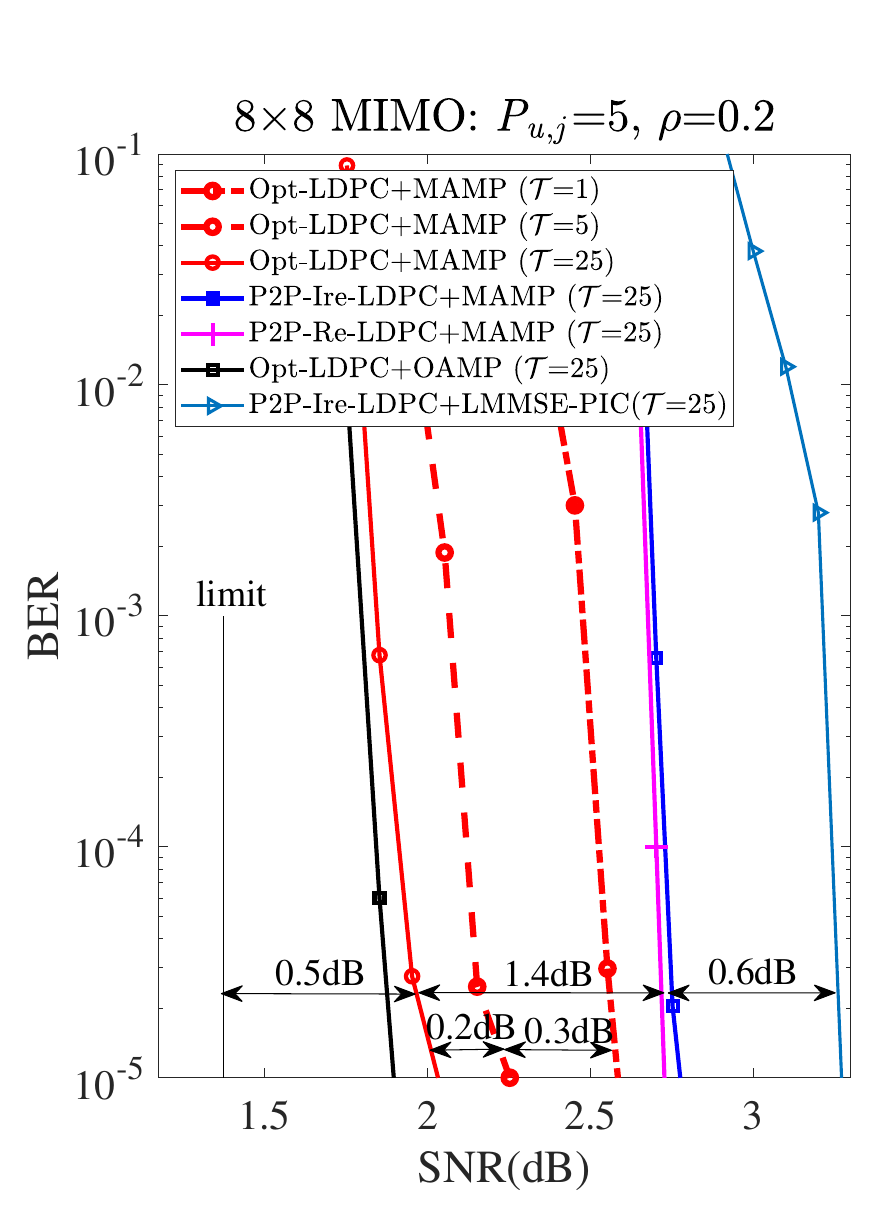}\label{fig:ber2b}
\end{minipage}
}\hspace{-0.5cm}
\subfigure[$4\times4 \& 8\times4$ MIMO: different $v$.]{
\begin{minipage}[t]{0.33\textwidth}
\centering
\includegraphics[width=0.7\textwidth]{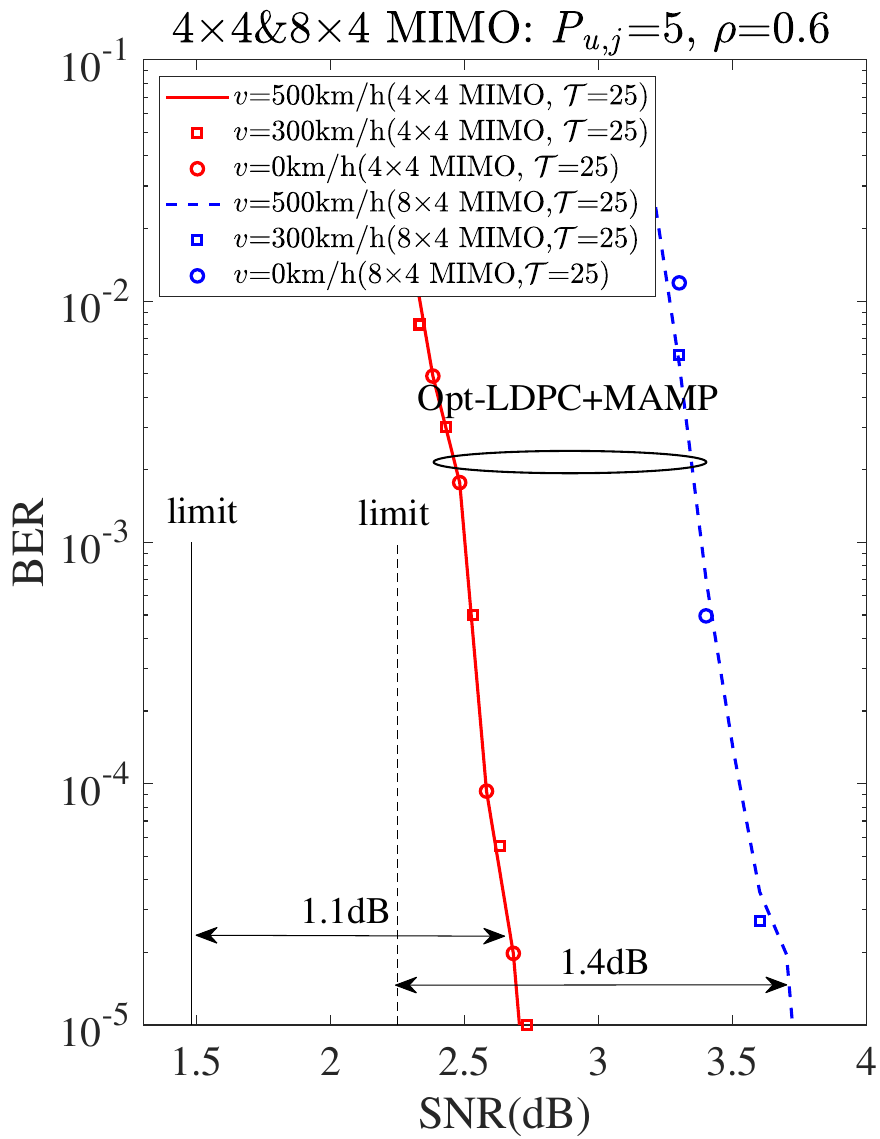}\label{fig:ber2c}
\end{minipage}
}
\caption{BER performance of MS-CD-MAMP and MS-CD-OAMP receivers with optimized LDPC codes in Table \ref{Tab:LDPCparam}, P2P regular (3,6) LDPC codes, P2P irregular LDPC codes in \cite{irrLDPC}, where $\mathcal{T}\in\{1, 5, 25\}$, $N=256$ for AFDM and OFDM and $K = 8$, $L = 32$ for OTFS in correlated MIMO channels with correlated parameters $\rho=\{0.2,0.6\}$ and number of multipaths $P_{u,j} \in\{2,5\}$.
For simplified notations, MS-CD-OAMP is denoted as OAMP and MS-CD-MAMP is denoted as MAMP.}\label{Fig:BER}
\end{figure*}

\subsection{BER Simulations and Comparisons of Finite Length Codes}

\emph{1})  \emph{BER comparison for OTFS/AFDM/OFDM:} Fig. \ref{fig:ber1a} shows the BER comparisons between OTFS, AFDM, and OFDM with the MS-CD-MAMP receiver and optimized LDPC codes presented in Table~\ref{Tab:LDPCparam}. We also provide the performance of well-designed P2P irregular LDPC codes as baselines, where the degree distributions of the well-designed P2P irregular LDPC code are $\lambda(X)=0.24426 x+0.25907 x^2+0.01054 x^3+ 0.05510 x^4+0.01455 x^7+0.01275 x^9+0.40373 x^{11}$ and $\mu(X)= 0.25475 x^6+0.73438 x^7+0.01087 x^8,$ whose rate $R_{\text {LDPC}} $\ is 0.5 and the decoding threshold is 0.18 dB away from the P2P-AWGN capacity\cite{irrLDPC}. The following results can be observed: 

    $\bullet$ For $\mathcal{T}=25$, the BER performance of OTFS, AFDM, and OFDM with optimized or P2P LDPC codes is almost the same, in which the performance of optimized LDPC codes is $1.8$ dB away from the limit at BER$=10^{-5}$ and have a $4.3$ dB gain over P2P LDPC codes. Meanwhile, the MS-CD-MAMP receiver can achieve the same BER performance as MS-CD-OAMP receiver when using the same optimized LDPC codes.
    
    $\bullet$ For $\mathcal{T}=1$, the BER performance of OTFS and AFDM is the same, but OFDM is slightly worse, with a loss of $0.4$ dB when applying P2P LDPC codes and $0.2$ dB when applying optimized LDPC codes. Meanwhile, The optimized LDPC codes can achieve about $4$~dB gains over P2P LDPC codes.
    
    This validates that OFDM can achieve performance close to OTFS and AFDM with the low-complexity MS-CD-MAMP receiver and optimal coding. As a result, to simplify the analysis, subsequent experiments are conducted by using OTFS as an example, and no further distinctions are made among these multicarrier modulations.
    
\emph{2})  \emph{BER comparison for different antenna correlation $\rho$:} 
Figs.~\ref{fig:ber1b},~\ref{fig:ber2a},~\ref{fig:ber2b} show the BER comparisons of the MS-CD-MAMP receiver with optimized LDPC codes in Table \ref{Tab:LDPCparam}, P2P regular (3,6) LDPC codes, and P2P irregular LDPC codes \cite{irrLDPC} in $8\times4$ and $8\times8$ MIMO, in which the antenna correlation $\rho \in\{0.2, 0.6\}$. Note that MS-CD-MAMP with optimized LDPC codes can achieve performance close to MS-CD-OAMP within about $0.1$~dB, while having about $1.1 \sim 4.7$ dB performance gains over MS-CD-MAMP with P2P regular and irregular LDPC codes. Meanwhile, the difference in BER performance of optimized LDPC codes with different antenna correlations is within 0.2 dB, confirming the robustness of the proposed scheme. 

\emph{3})  \emph{BER comparison for different number of multipath $P_{u,j}$:} Figs. \ref{fig:ber1a} and \ref{fig:ber1b} demonstrate the BER comparison of MS-CD-MAMP with optimized LDPC codes in $8\times 4$ MIMO, where the number of multipath $P_{u,j} \in \{2, 5\}$ and antenna correlation $\rho=0.6$. It is observed that MS-CD-MAMP with optimized LDPC codes can achieve about $4$~dB gains at $P_{u,j}=5$ over $P_{u,j}=2$ when BER=$10^{-5}$.  This also confirms that multicarrier modulation with MS-CD-MAMP and optimal coding scheme can exploit the multipath diversity gains for better performance.

\emph{4})  \emph{BER comparison for different time slots $\mathcal{T}$: } Figs.~\ref{fig:ber1c}, \ref{fig:ber2a}, and \ref{fig:ber2b} present the BER comparisons of the MS-CD-MAMP receiver and optimized LDPC codes with different $\mathcal{T}\in\{1, 5, 25\}$ in $4 \times 4$ and $8\times 8$ MIMO channels. Note that the performance of MS-CD-MAMP with optimized LDPC codes at $\mathcal{T} = 5$ and $\mathcal{T} = 1$ has performance losses of about $0.2 \sim 0.5$ dB and $0.6 \sim 0.8$~dB over $\mathcal{T} = 25$, respectively, but they still outperform MS-CD-MAMP with P2P regular and irregular LDPC codes at $\mathcal{T} = 25$. This also validates the importance of developing an optimal coding scheme for MIMO-multicarrier systems. \LL{
Meanwhile, compared with existing LMMSE-PIC\cite{LMMSE-PIC} with P2P irregular LDPC codes, the proposed MS-CD-MAMP receiver, employing both optimized and P2P irregular LDPC codes, achieves performance gains of approximately $1.5 \sim 2$ dB and $0.4 \sim 0.6$ dB in $8 \times 8$ MIMO multicarrier systems for $\rho = 0.6$ and $0.2$, respectively.
}
    
\emph{5}) \emph{BER comparison with different device velocities $v$:} Fig.~\ref{fig:ber2c} shows the BER comparisons of MS-CD-MAMP receiver and optimized LDPC codes with different device velocities $v \in \{0, 300, 500\}$ km/h in $4\times 4$ and $8\times 4$ MIMO channels. Note that  BER performances of MS-CD-MAMP with optimized LDPC codes for different $v$ are the same, which are $1.1\sim 1.4$~dB away from the associated limits. This is consistent with the achievable rate analysis in Fig.~\ref{Fig:MIMOspeed}, which indicates that the optimized LDPC codes are robust to the device velocity.

\LL{
\emph{6}) \emph{Spatial diversity gain with different receive antenna $U$:} By comparing Fig.~\ref{fig:ber1b} and Fig.~\ref{fig:ber2a}, it can be observed that when the number of transmit antennas $J=8$, increasing $M$ from $4$ to $8$ allows MS-CD-MAMP and MS-CD-OAMP with optimized LDPC codes to achieve roughly $1.7$~dB gain, with the corresponding theoretical limit improving by $0.7$ dB. This confirms that additional receive antennas provide extra spatial diversity gain.
}

\begin{figure}[t]
\centering 
\includegraphics[width=0.65\columnwidth]{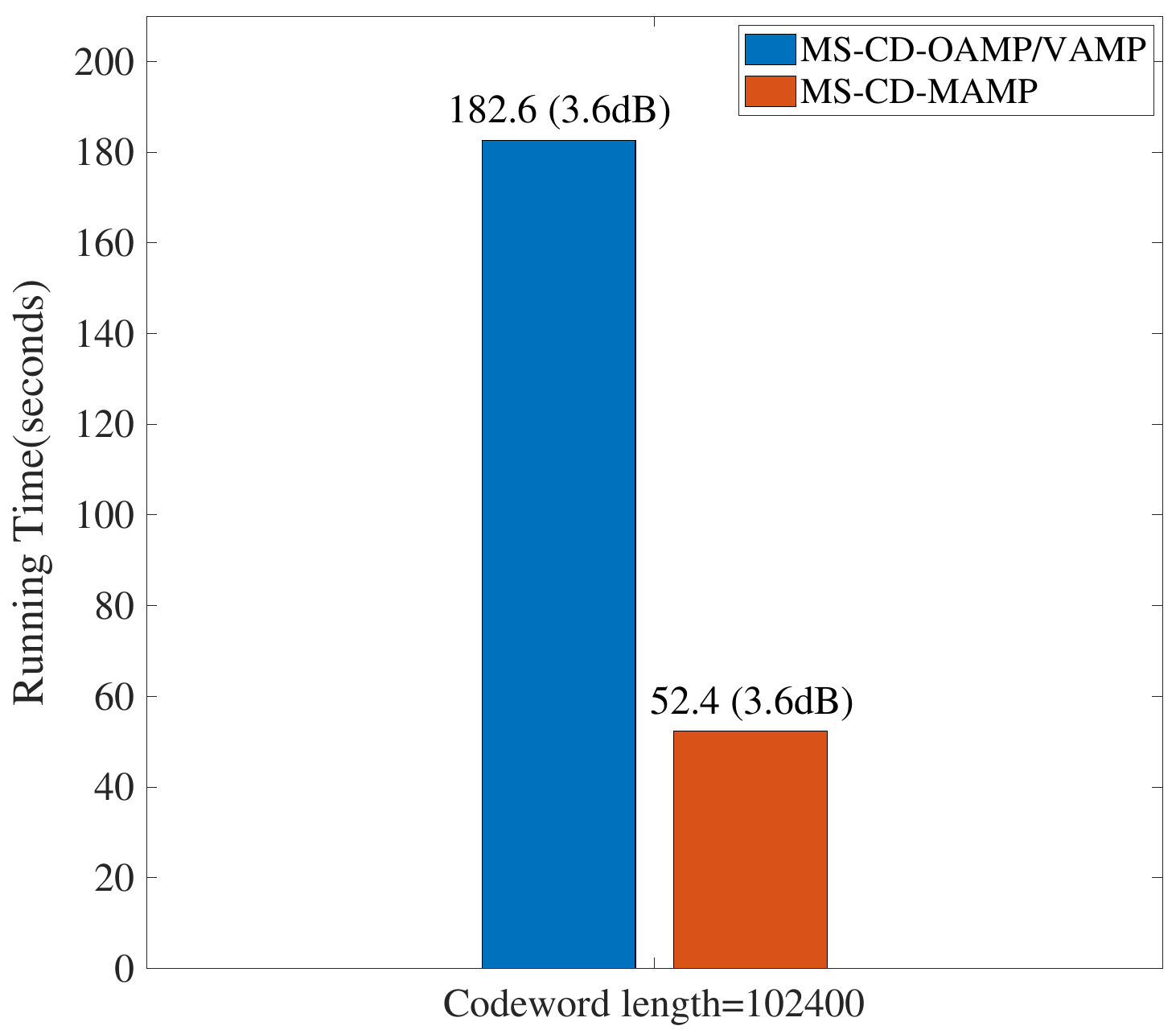}
 \caption{Running time comparison between MS-CD-MAMP and MS-CD-OAMP/VAMP receivers with optimized LDPC codes where $\mathcal{T}=25$, $N=256$ in correlated $8\times4$ MIMO channel with correlated parameters $\rho=0.6$ and number of multipaths $P_{u,j}=5$, $v=300$km/h.}\label{Fig:timecomplex}  
 \end{figure}

{
\subsection{Time Complexity of MS-CD-MAMP}\label{sec:timecomplex}
To highlight the advantage of the proposed MS-CD-MAMP receiver in terms of processing delay, we compare the time complexity of MS-CD-MAMP with that of MS-CD-OAMP/VAMP receivers in Fig.~\ref{Fig:timecomplex}. The results show that, at a BER of $10^{-5}$, the running time of MS-CD-MAMP is only approximately $28\%$ of that of MS-CD-OAMP/VAMP. The running time is obtained by Matlab 2024a on a PC with an AMD Ryzen 9 8945HS CPU and 32 GB of RAM.
}

\section{Conclusion}
This paper provides a low-complexity high-reliability MS-CD-MAMP receiver for coded MIMO-multicarrier systems, based on which the information-theoretic (i.e., achievable rate) limit analysis and the optimal coding principle are derived for arbitrary input distributions and various system configuration parameters. To achieve low-complexity signal recovery, multiple matched filters are employed for time-domain estimation to fully exploit the sparsity of the time-domain channels, resulting in a high-dimensional complex state evolution that is difficult to analyze directly. To address this difficulty, a simplified SISO VSE is proposed to analyze the achievable rate and derive the optimal coding principle. Numerical results show that MIMO-OFDM/OTFS/AFDM with MS-CD-MAMP and optimized LDPC codes can achieve the same achievable rate and approximately the same BER performance, which outperform those with well-designed P2P LDPC codes significantly. Furthermore, the proposed receiver and theoretical analysis can be applied to other advanced modulation schemes (e.g., IFDM), which is an interesting future work.

\appendix
\section*{Proof of Lemma~\ref{lem:VSE_MAMP}}\label{APP:VSE_MAMP}
We provide a detailed derivation of the VSE, which primarily relies on the fixed-point consistency between the SE of MS-CD-MAMP and MS-CD-OAMP/VAMP in Lemma \ref{lem:fixedpoint}, along with the SE and VSE functions of OAMP \cite{LeiOptOAMP}.{ Since both linear and nonlinear detections in the original SE of OAMP include orthogonal operations, it is unable to analyze the achievable rate directly using the I-MMSE lemma. To solve this difficulty, the VSE of OAMP was proposed in \cite{LeiOptOAMP}, i.e.,
\BS\label{Eqn:OAMP-VSE}
\begin{align}
\mathrm{LD:}\;\;\;\;\;\;\rho_{\ell} &= \eta^{\rm{O}}_{\rm{SE}}(v_{\ell}) = (v_{\ell})^{-1}-[\hat{\eta}_{\mr{SE}}^{-1}(v_{\ell})]^{-1},\label{Eqn:OAMP-VSE-LD}\\
\mathrm{NLD:}\;\;\;v_{{\ell}+1} &= \hat{\phi}^{\mathcal{C}}_{\rm{SE}}(\rho_{\ell}),\label{Eqn:OAMP-VSE-NLD}
\end{align}
\ES
where $\hat{\eta}_{\mr{SE}}(v_{\ell})= \tfrac{1}{N}{\rm{tr}}\{[snr\bf{H}^{\rm{H}}_t\bf{H}_t+v_{\ell}^{-1}\bf{I}]^{-1}\}$ denotes the MSE function of LMMSE detector, and $\hat{\eta}_{\mr{SE}}^{-1}(\cdot)$ denotes the inverse of $\hat{\eta}_{\mr{SE}}(\cdot)$.}

{Based on \eqref{Eqn:OAMP-VSE}, the VSE of the MS-CD-OAMP/VAMP receiver in MIMO multicarrier systems can be derived, where the received signals across $\mathcal{T}$ slots are independently processed by linear detection in time domain followed by averaging the estimation variances. That is, 
\BS\label{Eqn:VSE_MAMP2}
\begin{align}
    \!\!\!\!\!\mathrm{TDD}: \;\; & \rho^{\gamma}_{\ell}= \eta^{\rm{O}}_{\rm{SE}}(v^{\hat{\phi}}_{\ell}) = (v^{\hat{\phi}}_{\ell})^{-1}\!-\![\frac{1}{\mathcal{T}}\sum_{t=1}^{\mathcal{T}}\hat{\eta}_{t,\mr{SE}}^{-1}(v^{\hat{\phi}}_{\ell})]^{-1}, \label{Eqn:VSE-MLD2}\\
     \!\!\!\!\!\mathrm{NLD}:\;\; & v^{\hat{\phi}}_{\ell+1} = \hat{\phi}^{\mathcal{C}}_{\rm{SE}}(\rho^{\gamma}_{\ell}).\label{Eqn:VSE-NLD2}
\end{align}
\ES
}

\vspace{-0.2cm}
{Then, based on the VSE of MS-CD-OAMP/VAMP in \eqref{Eqn:VSE_MAMP2} and the Lemma \ref{lem:fixedpoint}, we assume that the VSE fixed point of MS-CD-OAMP/VAMP is $(\rho_*^{\mr{O}}, v_*^{\mr{O}})$. Similar to \cite{LeiOptOAMP}, based on the fixed-point equation equivalence of SE and VSE of MS-CD-OAMP/VAMP, we can derive that $\rho_*^{\mr{O}}=1/v_{*}^{\gamma,\rm{O}}$ and $v_*^{\mr{O}}=[(v_{*}^{\gamma,\rm{O}})^{-1}+(v_{*}^{\phi,\rm{O}})^{-1}]^{-1}$, where $(v_{*}^{\gamma,\rm{O}}, v_{*}^{\phi,\rm{O}})$ is the SE fixed point of MS-CD-OAMP/VAMP.
} 

{
For equivalent MS-CD-MAMP in Fig.~\ref{Fig:systemModel2}, assume the VSE fixed point is $(\rho_*^{\gamma}, v_*^{\hat{\phi}})$, where $\rho_*^{\gamma}=1/v^{\gamma}_{*}$. Due to the orthogonalization, we have 
$v^{\phi}_{*} =  [(v^{\hat{\phi}}_{*})^{-1} - (v^{\gamma}_{*})^{-1}]^{-1}$, i.e., $v^{\hat{\phi}}_{*} =  [(v^{\gamma}_{*})^{-1} + (v^{\phi}_{*})^{-1}]^{-1}$.
Meanwhile, based on Lemma~\ref{lem:fixedpoint}, for fixed $\hat{\phi}^{\mathcal{C}}_{\rm{SE}}(\cdot)$, MS-CD-MAMP and MS-CD-OAMP/VAMP converge to the same SE fixed point, i.e., $(v_{*}^{\gamma}, v_{*}^{\phi})=(v_{*}^{\gamma,\rm{O}}, v_{*}^{\phi,\rm{O}})$.
Therefore, MS-CD-MAMP and MS-CD-OAMP/VAMP have the same VSE fixed point, i.e., 
\BE\label{Eqn:vse_sf}
(\rho_*^{\gamma}, v_*^{\hat{\phi}})=(\rho_*^{\rm{O}}, v_*^{\rm{O}}). 
\EE
This indicates that the original SE and the VSE of MS-CD-OAMP/VAMP and MS-CD-MAMP converge to the same MSE performance, i.e., they share the same fixed point.
}

\begin{figure}[t]
	\centering 
	\includegraphics[width=0.65\columnwidth]{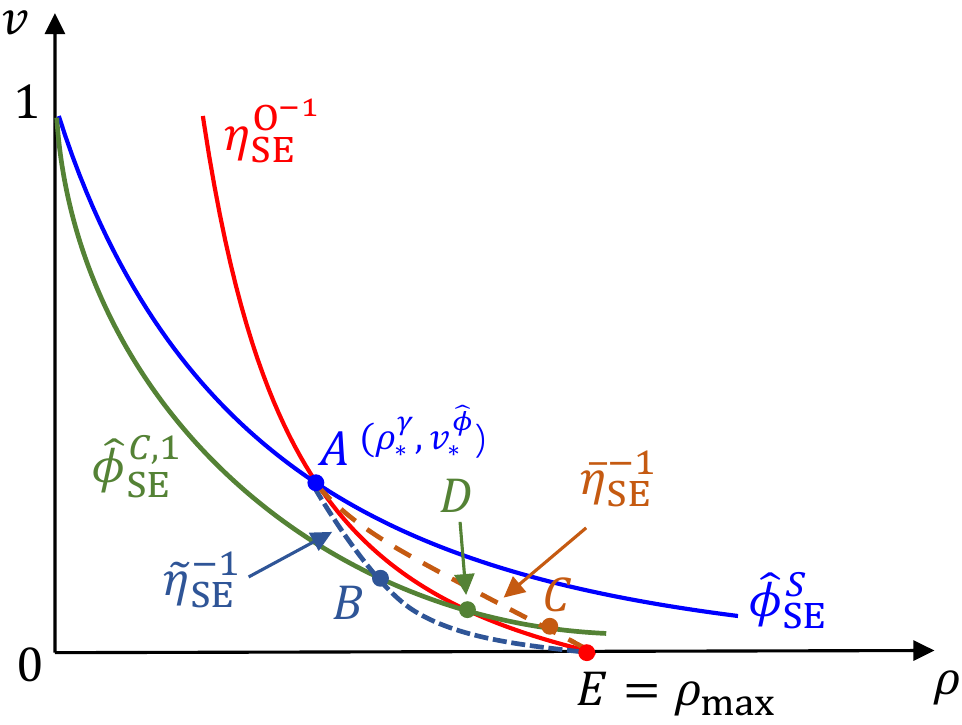}
    \caption{The VSE transfer functions for MS-CD-OAMP/VAMP and MS-CD-MAMP, where $\eta_{\mr{SE}}^{-1}(\cdot)$ is the inverse of $\eta_{\mr{SE}}(\cdot)$ and $\hat{\phi}^{\mathcal{S}}_{\rm{SE}}(\cdot)$ is the MMSE function of demodulation. MS-CD-MAMP and MS-CD-OAMP/VAMP have the same VSE fixed point~A. $\tilde{\eta}^{-1}_{\rm{SE}}(\cdot)$ and $\bar{\eta}^{-1}_{\rm{SE}}(\cdot)$ are two candidate inverse  of VSE transfer functions of MS-CD-MAMP's MLD. Given an MMSE function $\hat{\phi}^{\mathcal{C}}_{\rm{SE}}(\cdot)$ of decoder, its intersections with $\tilde{\eta}^{-1}_{\rm{SE}}(\cdot)$, $\bar{\eta}^{-1}_{\rm{SE}}(\cdot)$, and $\eta_{\mr{SE}}^{-1}(\cdot)$ are points B, C, and D, respectively.}\label{Fig:fixedpoint}  
\end{figure}

{
As shown in Fig.~\ref{Fig:fixedpoint}, assume that there is a unique VSE fixed point A $=(\rho_*^{\mr{O}}, v_*^{\mr{O}})$ between $\eta^{{\mr{O}}^{-1}}_{\rm{SE}}(\cdot)$ and $\hat{\phi}^{\mathcal{S}}_{\rm{SE}}(\cdot)$ in \eqref{Eqn:VSE_MAMP}, where $\hat{\phi}^{\mathcal{S}}_{\rm{SE}}(\cdot)$ is the MMSE function of demodulation. 
Based on \eqref{Eqn:vse_sf}, the VSE fixed point between $\eta^{-1}_{\rm{SE}}(\cdot)$ and $\hat{\phi}^{\mathcal{S}}_{\rm{SE}}(\cdot)$ for MS-CD-MAMP is also point A~$=(\rho_*^{\gamma}, v_*^{\hat{\phi}})$. When $0\le\rho^{\gamma}<\rho^{\gamma}_{*}$, MS-CD-MAMP can iterate with $\hat{\phi}^{\mathcal{S}}_{\rm{SE}}(\rho^{\gamma})<\eta^{-1}_{\rm{SE}}(\rho^{\gamma})$. Furthermore, $\hat{\phi}^{\mathcal{C}}_{\rm{SE}}(\rho^{\gamma})<\hat{\phi}^{\mathcal{S}}_{\rm{SE}}(\rho^{\gamma})$ is obtained because of the coding gain.
Therefore, $\hat{\phi}^{\mathcal{C}}_{\rm{SE}}(\rho^{\gamma})<\hat{\phi}^{\mathcal{S}}_{\rm{SE}}(\rho^{\gamma})<\eta^{-1}_{\rm{SE}}(\rho^{\gamma})$, i.e., $\hat{\phi}^{\mathcal{C}}_{\rm{SE}}(\rho^{\gamma})$ is limited to $\hat{\phi}^{\mathcal{S}}_{\rm{SE}}(\rho^{\gamma})$, and it is reasonable to ignore the impact of the specific expression of $\eta^{-1}_{\rm{SE}}(\rho^{\gamma})$ on $\hat{\phi}^{\mathcal{C}}_{\rm{SE}}(\rho^{\gamma})$.
Therefore, we only focus on the expression of $\eta_{\rm{SE}}^{-1}(\rho^{\gamma})$ for $\rho^{\gamma}\ge\rho^{\gamma}_{*}$.
}

{
Given $\forall \rho_1,\rho_2 \in[\rho^{\gamma}_{*},\rho_{\rm max}]$ with $\rho_1<\rho_2$, assume $\eta_{\rm{SE}}^{-1}(\rho^{\gamma})>\eta^{{\rm{O}}^{-1}}_{\rm{SE}}(\rho^{\gamma})$ for $\rho^{\gamma}\in [ \rho_1,\rho_2]$. Given an MMSE function $\hat{\phi}^{\mathcal{C},1}_{\rm{SE}}(\cdot)$ of decoder, due to the coding gain, $\hat{\phi}^{\mathcal{C},1}_{\rm{SE}}(\cdot)<\hat{\phi}^{\mathcal{S}}_{\rm{SE}}(\cdot)$ is obtained. Since MMSE function is monotone decreasing, $\hat{\phi}^{\mathcal{C},1}_{\rm{SE}}(\rho^{\gamma})$ has two different VSE fixed point with $\eta_{\rm{SE}}^{-1}(\rho^{\gamma})$ and $\eta^{\rm{O}^{-1}}_{\rm{SE}}(\rho^{\gamma})$, which contradicts \eqref{Eqn:vse_sf}.
Similarly, assume $\eta_{\rm{SE}}^{-1}(\rho^{\gamma})<\eta^{\rm{O}^{-1}}_{\rm{SE}}(\rho^{\gamma})$ for $\rho^{\gamma}\in [ \rho_1,\rho_2]$. There are still two different VSE fixed point with $\eta_{\rm{SE}}^{-1}(\rho^{\gamma})$ and $\eta^{\rm{O}^{-1}}_{\rm{SE}}(\rho^{\gamma})$, which contradicts \eqref{Eqn:vse_sf}. As a result, due to the arbitrariness of $\rho_1$ and $\rho_2$, $\eta_{\rm{SE}}^{-1}(\rho^{\gamma})=\eta^{\rm{O}^{-1}}_{\rm{SE}}(\rho^{\gamma})$ (i.e., $\eta_{\rm{SE}}(\rho^{\gamma})=\eta^{\rm{O}}_{\rm{SE}}(\rho^{\gamma})$) for $\rho^{\gamma}\in[\rho^{\gamma}_{*},\rho_{\rm max}]$ is proved. Since the specific expression of $\eta_{\rm{SE}}(\rho^{\gamma})$ for $\rho^{\gamma}\in[0,\rho^{\gamma}_{*})$ is negligible, we let $\eta_{\rm{SE}}(\rho^{\gamma})=\eta^{\rm{O}}_{\rm{SE}}(\rho^{\gamma})$ with $\rho^{\gamma}\in[0,\rho_{\rm max}]$ for simplicity.
Therefore, the VSE of MS-CD-MAMP is derived, as presented in Lemma \ref{lem:VSE_MAMP}.
}

{
Furthermore, to illustrate this proof intuitively, letting $\rho_1=\rho^{\gamma}_{*}$ and $\rho_2=\rho_{\rm max}$, two candidate functions ${\tilde{\eta}}^{-1}_{\rm{SE}}(\cdot)$ and ${\bar{\eta}}^{-1}_{\rm{SE}}(\cdot)$ of $\eta^{-1}_{\rm{SE}}(\cdot)$ are obtained in Fig.~\ref{Fig:fixedpoint}. The intersections of $\hat{\phi}^{\mathcal{C},1}_{\rm{SE}}(\cdot)$ with ${\tilde{\eta}}^{-1}_{\rm{SE}}(\cdot)$, ${\bar{\eta}}^{-1}_{\rm{SE}}(\cdot)$, and $\eta^{\rm{O}^{-1}}_{\rm{SE}}(\cdot)$ are point B, C, and D, respectively, which contradicts \eqref{Eqn:vse_sf}. Therefore, we can derive that $\eta_{\rm{SE}}(\cdot)=\eta^{\rm{O}}_{\rm{SE}}(\cdot)$.
}

\bibliographystyle{IEEEtran}
\bibliography{manuscript}

\end{document}